\documentclass{jfm}
\usepackage{graphicx}
\usepackage{epstopdf, epsfig}
\usepackage{natbib}
\usepackage{graphics}
\usepackage{amsmath}
\usepackage{amssymb}
\usepackage{mathrsfs}
\usepackage{wasysym}
\usepackage{caption}
\usepackage{subcaption}
\usepackage[hidelinks]{hyperref}
\pdfoutput=1

\shorttitle{SOC in strongly stratified shear flows}
\shortauthor{H. Salehipour, W. R. Peltier and C. P. Caulfield}

\title{Self-organized criticality of turbulence in strongly stratified mixing layers}

\author{Hesam Salehipour\aff{1,2}
  \corresp{\email{h.salehipour@utoronto.ca}},
  W. R. Peltier\aff{1}
 \and C. P. Caulfield\aff{3,4}}

\affiliation{\aff{1}Dept. of Physics, University of Toronto, Toronto, ON M5S 1A7, Canada
\aff{2}Autodesk Research, MaRS Discovery District, 661 University Ave, Toronto, ON M5G 1M1, Canada
\aff{3}BP Institute, University of Cambridge, Madingley Road, Cambridge CB3 0EZ, UK
\aff{4}Department of Applied Mathematics and Theoretical Physics, University of Cambridge, Centre for Mathematical Sciences, Wilberforce Road, Cambridge CB3 0WA, UK}

\begin{document}

\maketitle

\begin{abstract}
Motivated by the importance of stratified shear flows in geophysical and environmental circumstances, we characterize their energetics, mixing and spectral behavior through a series of direct numerical simulations of turbulence generated by Holmboe wave instability (HWI) under various initial conditions. We focus on circumstances where the stratification is sufficiently `strong' so that HWI is the dominant primary instability of the flow. Our numerical findings demonstrate the emergence of self-organised criticality (SOC) that is manifest as an adjustment of an appropriately defined gradient Richardson number, $Ri_g$, associated with the horizontally-averaged mean flow, in such a way that it is continuously attracted towards a critical value of $Ri_g \sim 1/4$. This self-organization occurs through a continuously reinforced localisation of the `scouring' motions (i.e. `avalanches') that are characteristic of the turbulence induced by the break down of Holmboe wave instabilities and are developed on the upper and lower flanks of the sharply localized density interface, embedded within a much more diffuse shear layer. These localised `avalanches' are also found to exhibit the expected scale invariant
characteristics. From an energetics perspective, the emergence of SOC is expressed in the form of a long-lived turbulent flow that remains in a `quasi-equilibrium' state for an extended period of time. Most importantly, the irreversible mixing that results from such self-organised behavior appears to be characterized generically by a universal cumulative turbulent flux coefficient of $\Gamma_c \sim 0.2$ \emph{only} for turbulent flows engendered by Holmboe wave instability. The existence of this self-organised critical state corroborates the original physical arguments associated with self-regulation of stratified turbulent flows as involving a `kind of equilibrium' as described by \cite{Turner_1979}. 
\end{abstract}


\section{Introduction} 
\label{sec:intro} 

Over a century of study of stratified shear flows, as pioneered by G.~I.~ \citet{Taylor_1915}, have yet left a number of important questions unanswered regarding the mysterious properties of such geophysically ubiquitous flows; from their mixing properties \citep{Linden_1979, PC03, ivey08, Gregg_etal_2018_review} to their anisotropic layered `pancake' structures \citep{Lilly_1983, Lindborg_2006}. In particular, the properties of stratified turbulence under the influence of \emph{strong} stratification are especially not well-understood,
though of course care must be taken in the definition and appreciation of what precisely is meant by `strong'. 

Under what he referred to as \emph{`very stable'} conditions, \citet{Turner_1979} has suggested (based on the Monin-Obukhov similarity theory) that in wall-bounded shear flows, the wall distance becomes irrelevant and stratified layers characterized by a constant flux are retained. \citeauthor{Turner_1979} has further argued that the flow in this strong stratification limit is in \emph{`a kind of equilibrium, self-regulated'} state; this being a state in which flow quantities such as the gradient Richardson number, $Ri_g$ and the flux Richardson number, $Ri_f$ (an approximation to the mixing efficiency) are internally regulated.

The local strength of the stabilizing density stratification relative to the local destabilizing influence of the velocity shear in stratified flows, whether or not they are turbulent, may be formally characterized by the gradient Richardson number, $Ri_g$, which may be defined as
\begin{equation}
 Ri_g(z,t) = \frac{N^2}{S^2} = {\left(-g/\rho_r\right) \partial\overline{\rho}/\partial z \over \left(\partial\overline{u}/\partial z\right)^2}
 \label{eq:Rig}
\end{equation}
where $\overline{\rho}(z,t)$ and $\overline{u}(z,t)$ denote respectively the horizontally-averaged and (generally) time-dependent vertical profiles of mean flow density and velocity, $g$ is the gravitational acceleration and $\rho_r$ is a hydrostatic reference density. $Ri_g$ usually exhibits significant variation with both large and small local values. Hence, it may be argued that it is more appropriate to classify a particular flow as being `strongly' stratified in terms of a `bulk' Richardson number $Ri_b$, defined as
\begin{equation}
  Ri_b=\frac{g \rho_0 d}{\rho_r U_0^2},
  \label{eq:Rib}
\end{equation}
in which $\rho_0$ and $U_0$ represent characteristic density and velocity changes across a characteristic length $d$ whose precise definitions depend on the flow geometry. Notice that it is naturally possible that a flow with a high value of $Ri_b$ still has non-trivial
regions of the flow where $Ri_g(z,t)$ is close to zero for significant periods of time.

Either in a bulk-averaged sense or as a point-wise value, in the literature, there are often connections drawn with the critical value of $Ri^{crit}_g = 1/4$ based on the inviscid linear theory of Miles and Howard \citep{Miles61,Howard61} who first demonstrated that the necessary condition for linear instability of a two-dimensional inviscid and steady stably stratified non-turbulent parallel flow is that $Ri_g < 1/4$ somewhere within the flow. Indeed, such connections have been made, even when the underlying assumptions of the Miles-Howard theoretical analyses categorically do not apply. It is perhaps not surprising that even in turbulent flows, relatively `small' values (i.e. close to $1/4$) of a Richardson number should emerge, as such values can be loosely thought of as being characteristic of stratified flows where the stratification has a weak but non-trivial effect on the flow's evolution. For example, a `stationary' characteristic value of $Ri_g \sim 0.21$ (largely independent of both time and the direction normal to the mean flow) has been reported on the basis of direct numerical simulations of the turbulence generated by stationary homogeneous stratified shear flows  \citep{Rohr_1988b, Holt_1992, Shih_2000}. More recently, a similar stationary value of $Ri_g$ has also been found in simulations of stratified plane Couette flow for sufficiently strong stratifications and sufficiently high Reynolds numbers \citep{ZTC17}. It is postulated in \cite{ZTC17} that such a characteristic value of $Ri_g$ might be associated with the existence of a turbulently balanced equilibrium state (according to the classic Monin-Obukhov similarity theory), and that the proximity of this characteristic value to the Miles-Howard criterion might be `fortuitous'. Furthermore, in the context of oceanic observations, $Ri_g\sim 1/4$ has been reported as a measured characteristic of equatorial undercurrent turbulence and this has been attributed to the attraction of the turbulence to a state of `marginal' instability \citep{Smyth_etal_2013_marginal, Thorpe_Liu_2009}. Similarly, bulk Richardson numbers $Ri_b \simeq 0.3$ have been observed throughout the entire length of the Burlington Ship Canal connecting Hamilton Harbour and Lake Ontario (see \cite{Lawrence_etal_2004}).

The above mentioned, largely physically-based and qualitative arguments presented by \cite{Turner_1979} predate but appear to be deeply connected to the more general concept of \emph{self-organized criticality} (SOC), originally proposed by \cite{BTW} in the context of dynamical systems (see \cite{aschwanden2016} for a recent review and \cite{SOC_book_2012} for a more in-depth introduction to the SOC literature). To set the stage for a discussion of the applicability of SOC to the understanding of the turbulent flows that are of interest in the present context it is useful to briefly reprise the basic characteristics of the sandpile `thought experiment' in terms of which the basic idea is usually discussed. Consider a sandpile that is formed by slowly dropping grains of sand onto a flat surface. The sandpile slope gradually increases to a critical value at which point the system is `marginally stable' and beyond which further addition of sand grains will locally destabilise the pile through the occurrence of local `avalanches' until the critical marginal state is restored. SOC has been observed in many \emph{slowly-driven, non-equilibrium} systems which involve spatiotemporal complexity that evolves, without external tuning, into a \emph{scale-invariant} state \citep{SOC_book_2012}.

SOC has been shown to be characteristic of a wide range of physical systems although it remains relatively unexplored in the context of geophysical fluid turbulence. One
exception to this concerns the thermal turbulence produced in a variant of the classical Benard problem of thermal convection, which involves a highly unstable layer of fluid that
is heated from below and cooled from above and which also undergoes an endothermic phase transition at a particular depth within the layer \citep{Solheim_Peltier_1994}. The thermal turbulence in this flow is characterized by episodic transitions between layered flow with strongly inhibited mass flux across the phase transition interface and intense  convective mixing throughout the domain. During the layered condition, a strong internal thermal boundary layer is established across the phase transition interface which becomes increasingly  well developed as the layered flow  evolves and eventually collapses through local convective instability of this boundary layer. The analogy here to the critical angle of repose of the sandpile is with the critical Rayleigh number associated with this internal boundary layer instability. Indeed, it was established by \cite{butler1997} that the avalanches of mass across this internal interface continuously restore the system to its critical state and that these avalanches exhibit self-similar scaling, a necessary characteristic of SOC behavior. A more recent example of an appeal to SOC and the sandpile analogy to understand the observed characteristics of a turbulent geophysical fluid flow has been that provided by \cite{Smyth_etal_2013_marginal} and \cite{Smyth_etal_2013_JPO}, who described observations of equatorial undercurrent turbulence as being in an apparent state of `marginal instability'. Their appeal to the SOC analogy was in the context of density stratified turbulent flows forced by vertical shear of the horizontal undercurrent velocity. Our interest in the present paper is in investigating whether the SOC concept can be applied to an unforced stratified shear flow, far from boundaries, particularly when the stratification is `strong', in a sense we will now make precise.


There is a particular class of `strongly stratified' shear flows which we wish to consider here. The stratification has two key characteristics of `strength'. First, there must be significant variation in $Ri_g(z,t)$, with the existence of a relatively thin `interface' of substantially enhanced density gradient (and hence `strong' local stratification neighboured by weaker stratification) embedded within a relatively deeper shear layer. Second, the overall bulk stratification (quantified by $Ri_b$) must be sufficiently high so that, in combination with the presence of significant vertical variation in $Ri_g$, the flow is
likely to become unstable due to a normal mode instability known as the Holmboe wave instability (HWI) \citep{Holmboe}. This instability can be interpreted as arising due to a resonant interaction of a vorticity wave (or a Rayleigh wave) localized at the edge of the shear layer, and an interfacial gravity wave (see \eg \cite{Caulfield_1994, Baines_1994, Guha_Lawrence_2014}), localized where the density gradient is relatively large. HWI, therefore, is an instability which relies inherently on the presence of (relatively strong) stratification
in this very specific sense. On the other hand, the classic `Kelvin-Helmholtz instability' (KHI), arising (in the non-singular case of a finite depth shear layer) due to a resonant interaction between two vorticity waves localized at either edge of the shear layer, is monotonically stabilized by increasing stratification, and indeed does not grow under arbitrarily large levels of stratification, whereas provided $Ri_g$ varies in the above-mentioned way, HWI is predicted to occur (for some range of wavenumbers) for arbitrarily high values of $Ri_b$, and thus for `strongly' stratified shear flows.

Considering idealized symmetric two-layer configurations with a relatively `sharp' density interface embedded at the midpoint of the shear layer, $Ri_g$ becomes less than $1/4$ above and below the interface, (thus not violating the Miles-Howard criterion) leading to the emergence of HWI, manifest at finite amplitude as two counter-propagating cusped waves on the interface, induced by counter-propagating vortices above and below the interface (see for example early experimental observations of \cite{macagno1961} and \cite{thorpe1968}, and early numerical observations of \cite{Smyth_1988_Holmboe}, and for a more detailed historical discussion \cite{lefauve2018}). Provided that the Reynolds number is sufficiently large, HWI is itself subject to a host of secondary instabilities which mediate the transition to a fully turbulent state that supports a $-5/3$ power-law kinetic energy spectrum for horizontal scales in excess of the Ozmidov scale (as defined below) \citep{SCP15}.
Although many flows, both in nature and the laboratory,  exhibit some asymmetry with the sharp density interface 
offset from the midpoint of the shear layer, leading to the favouring of one or other of the cusped waves
(as theoretically predicted by \cite{Lawrence_1991} and experimentally observed by, among others \cite{Zhu_Lawrence_2001}) we consider symmetric flows here for simplicity. The ensuing turbulence above and below the initial density interface associated with the break down of the primary instabilities appears to be largely decoupled from the turbulence on the other side of the interface, and so it is natural to consider a symmetric flow to 
obtain as much turbulent data as possible.

\begin{figure}
  \centerline{\includegraphics[width=.8\textwidth]{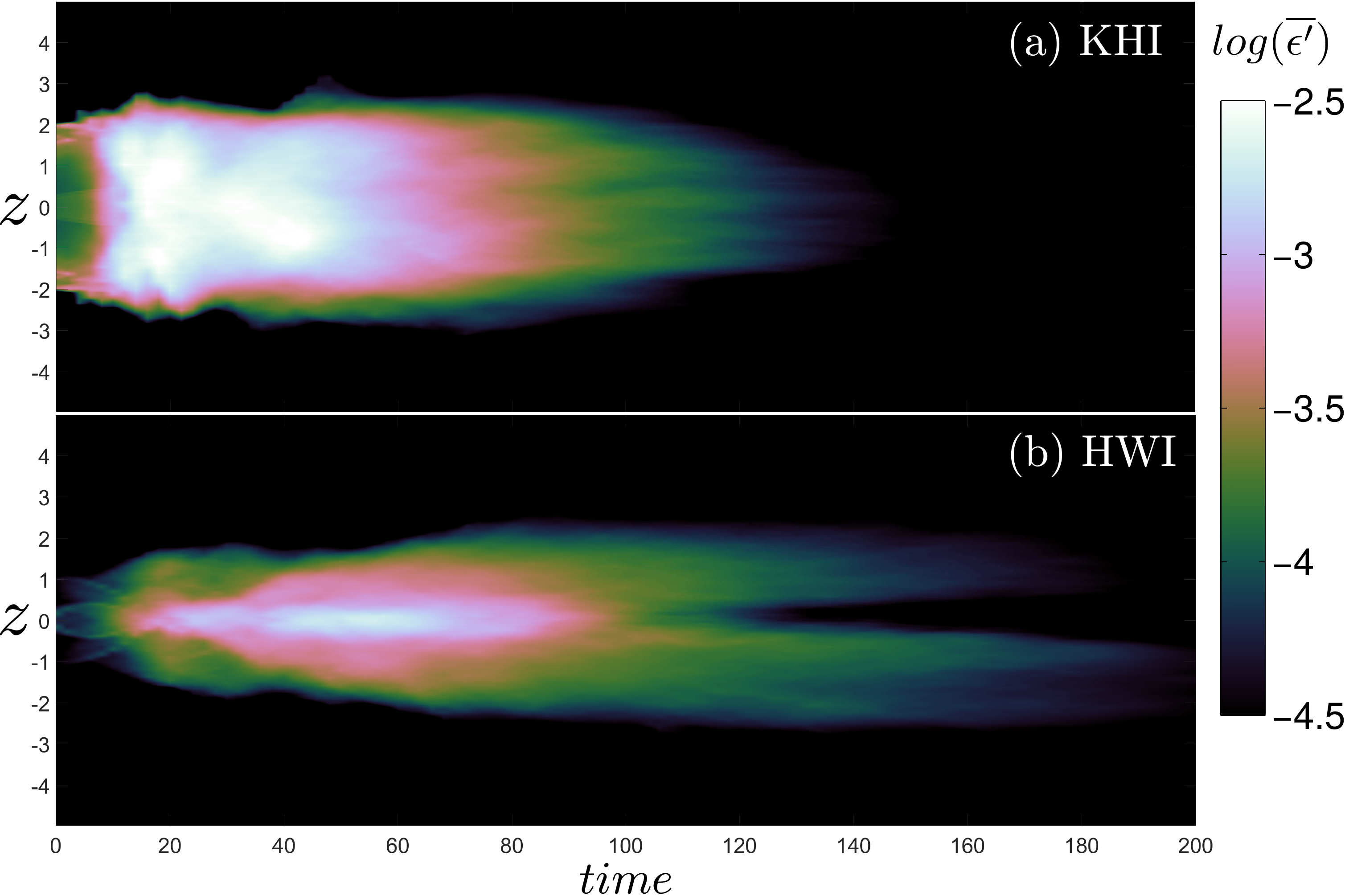}}
  \caption{The variation with time of the vertical structure of the horizontally-averaged turbulent dissipation $\overline{\epsilon'}(z,t)$ (as defined in \eqref{eq:epsilon_z}) due to: (a) KHI and (b) HWI. Vertical scales are non-dimensionalized with $d$, the initial shear layer half depth, while  time is non-dimensionalized with the advective time scale $d/U_0$ where  $U_0 = \Delta U/2$ is  half the velocity difference across the shear layer (see \ref{eq:tanhprof}). Only times subsequent to $t_{2d}$ (the time when the spanwise-averaged perturbation is maximised) are plotted (see \cite{SPM15} for further details).}
  \label{fig:longevity}
\end{figure}

In \cite{SCP15}, we compared the longevity of the turbulence induced by KHI and HWI and found that the turbulence induced by HWI is noticeably longer-lived than that induced by KHI: in some sense the KHI `flares' then rapidly dies, while HWI `burns', decaying more slowly. The difference in their respective life-spans of turbulence can be attributed to the mechanics and localization of mixing that is specific to each primary instability mechanism. While mixing occurs most prominently in KHI due to a vigorous `overturning' of the relatively weak density interface by the primary KH `billows', HWI-induced mixing is localized at the flanks of the relatively strong or sharp interface and is characterized by `scouring' motions, associated with the turbulence arising from the break down of the primary counter-propagating vortices. Figure \ref{fig:longevity} illustrates these two key qualitative differences (of localization and duration of turbulence) using the results associated with two direct numerical simulations that will be further analysed in what follows. The temporal evolution of vertical profiles of horizontally-averaged turbulent dissipation $\overline{\epsilon'}(z,t)$ is illustrated in this figure, where $\overline{\epsilon'}(z,t)$ is defined as
\begin{equation}
  \overline{\epsilon'} (z,t) = 2\nu \overline{ s'_{ij}s'_{ij}},
  \label{eq:epsilon_z}
\end{equation}
in which $\nu$ is the kinematic viscosity, $s'_{ij} = \left( \partial u'_i / \partial x_j + \partial u'_j / \partial x_i \right)/2 $ is the disturbance strain rate tensor and  $\boldsymbol{u}' = (u',v',w')$ represents the perturbation (away from the horizontally-averaged mean) velocity field. The relatively slow evolution of HWI, driven by localized scouring motions that leads to a long-lived turbulent state might be contrasted with the relatively sudden burst of KHI into turbulence. This key characteristic of HWI is in accord with that required for a complex system to be self-organized towards a critical point, and thus we may conjecture \emph{a priori} that flows unstable to HWI, rather than to KHI, are conceivably candidate flows that might support SOC.

The goal of the present paper is to investigate the validity of this conjecture especially in connection to the earlier ideas for `strongly stratified' flows envisioned by \citeauthor{Turner_1979}. A related fundamental question is whether the 
proximity of the mean turbulent flow characteristics to a critical value of $Ri_g \sim 1/4$ in both observations and numerical simulations, as discussed above, is fortuitous (as postulated by \cite{ZTC17}) or perhaps speaks to a more universal behavior that is inherently connected to internal regulation of the flow dynamics. A key concept is that the classification of a flow as being `strongly stratified' must be done with great care. As described above, we refer to a shear flow as being `strongly stratified' in a very particular sense corresponding to when the initial velocity and 
density distributions are such that there is an interfacial region of locally high $Ri_g$, and also the overall value of $Ri_b$ is sufficiently high (and indeed maybe greater than $1/4$) so that the flow is susceptible to primary HWI. We find that such flows
%
with relatively large values of $Ri_b$ can still have  $Ri_g \sim 1/4$ over much of the shear layer while the flow is turbulent. Therefore, while \emph{globally} the flow might be thought to be strongly stratified, \emph{locally} the stratification is not sufficiently strong to suppress vigorous turbulence, and indeed as we will demonstrate the flow locally self-organizes to a critical stratification such that $Ri_g \sim 1/4$.

Thus, we seek herein to investigate quantitatively whether SOC emerges under strongly stratified conditions (in the above sense) following the turbulent break down of HWI. For this purpose we study a series of turbulent flows induced by HWI and aim to understand and characterize (i) the development of a `kind of equilibrium' (as proposed by Turner) in these flows, (ii) the self-regulation of $Ri_g$ throughout the flow evolution, (iii) the self-regulation of turbulence energetics and in particular self-regulation of a key (and controversial) measure of the `efficiency' of mixing, and finally (iv) the emergence of scale-invariance. To explore these issues, the remainder of this paper is organised as follows. In \S\ref{sec:method}, we briefly describe a series of direct numerical simulations (DNS) of HWI and introduce the quantities required for characterization of the mean flow and energetics of the stratified turbulence produced by flow transition. We discuss our results in \S\ref{sec:results} while in \S\ref{sec:summary} we summarize our conclusions.

\section{Methodology}
\label{sec:method}
%
We consider the canonical choice of hyperbolic tangent initial velocity and density distributions:
\begin{equation}
  \overline{u}(z,0)= U_0 \tanh \left (\frac{z}{d} \right ), 
  \quad
  \overline{\rho} (z,0) = \rho_0 \left[ 1 -  \tanh   \left (\frac{z}{\delta} \right ) \right], 
  \label{eq:tanhprof}
\end{equation}
in the Boussinesq approximation (with a linear equation of state) such that $\rho_0 \ll \rho_r$ (note that here density represents departures from a hydrostatic state associated with $\rho_r$). In \eqref{eq:tanhprof} $U_0=\Delta U/2$ denotes half the total velocity difference across the shear layer thickness of total depth $2d$. Similarly $\rho_0 = \Delta \rho/2$ is half the total density difference across the density layer thickness of total depth $2\delta$. Besides the bulk Richardson number, $Ri_b$, (as defined previously in \eqref{eq:Rib}, sometimes also denoted as $J$) there are three additional important non-dimensional parameters that govern the non-dimensional Boussinesq equations: the (initial) Reynolds number $Re_0$; the Prandtl number $Pr$; and the initial scale ratio $R$, defined altogether as:
\begin{equation}
Ri_b=\frac{g \rho_0 d}{\rho_r U_0^2},
\quad
Re_0= \frac{U_0 d}{\nu}, 
\quad 
Pr = \frac{\nu}{\kappa},
\quad
R=\frac{d}{\delta}, 
\label{eq:params}
\end{equation}
where $\kappa$ is the value of molecular thermal diffusivity. For comparison with other numerical, experimental and observational analyses, it is important to note that $Re_0$ is defined using \emph{half} the total shear layer depth and \emph{half} the total velocity difference across the shear layer. 
In particular, it is important to appreciate that the relatively long-lived turbulent state considered here for flows primarily susceptible 
to HWI relies critically on the Reynolds number of the flow. As discussed in more detail in \cite{SCP15}, the 
behaviour of flows at $Re=500$ is qualitatively different from the truly turbulent flows which develop for $Re_0 > 4000$. In this paper, we choose $Re_0=6000$, and so the inherently turbulent flows are qualitatively different from previous studies such as those of \cite{Smyth_Winters_2003}, where the Reynolds number is twenty times smaller, namely $Re_0=300$ (using our convention), and so the flow is much more strongly affected by viscosity.

For our choice of density and velocity profiles defined in \eqref{eq:tanhprof}, the initial value of the gradient Richardson number at the midpoint of the shear layer where $z=0$, $Ri_g(0,0)$ (denoted by $Ri_g(0)$ for brevity) is related to $Ri_b$ and $R$ through
\begin{equation}
 Ri_g(0)= Ri_b R.
\end{equation}
Crucially, as noted by \cite{Smyth_1988_Holmboe}, for sufficiently `sharp' density interfaces with $R > 2$, $Ri_g(0)$ is the \emph{maximum} value of $Ri_g(z,0)$ across all values of $z$, and therefore flows with high $R$ can be susceptible to HWI for arbitrarily large values of $Ri_b$, while conversely for flows with $O(1)$ values of $R$, the primary KHI is suppressed when $Ri_g(0)\sim Ri_b > 1/4$.

We employ direct numerical simulation (DNS) of the three-dimensional governing equations under the Boussinesq approximation using the spectral element solver, Nek5000 \citep{Fischer_1997}. The flow is assumed to be periodic in the horizontal plane while the top and bottom boundaries are free-slip and impermeable for the velocity fields and insulating for the density field. For details of the numerical methodology, boundary conditions and simulation setup, the interested reader is referred to \cite{SPM15,SCP15}. 

Table \ref{tab:siminfo} lists the initial conditions for various cases of HWI as well as a single case of KHI that will be discussed in this paper. Note that except for the KHI simulation in table \ref{tab:siminfo} which has been reported previously \citep{SP15}, the remaining  DNS analyses associated with HWI have not been described previously in the literature. Note also that the employed KHI simulation only differs from case R3-J016 in terms of the initial thickness ratio of the shear and stratified regions (i.e. $R=1$ for KHI and $R=\sqrt{8}$ for HWI). Furthermore, in our discussion below three characteristic times during the flow evolution will be employed. The time $t_{2d}$ represents the time associated with the maximum amplitude of the spanwise-averaged perturbation, while $t_{3d}$ denotes the characteristic time of the maximum amplitude of the inherently three-dimensional deviation from this perturbation (as defined precisely in \cite{SPM15}). This time also may be thought of as an identifier of the onset of a fully turbulent stage in the flow evolution. Finally, the time $t_{rl}$ marks the onset of a re-laminarization stage, defined here as the time when the buoyancy Reynolds number (also sometimes referred to as the turbulence intensity parameter) decreases to $Re_b \leq 7$. Here $Re_b$ is defined as
\begin{equation}
  Re_b = \frac{\langle \bar{\epsilon'} \rangle}{\nu \langle N^2 \rangle},
  \label{eq:Reb}
\end{equation}
using the buoyancy frequency as defined in \eqref{eq:Rig}. This numerical value is consistent with the value proposed by \cite{ivey08} for stratified turbulence being in the `molecular'
regime. Note that everywhere in this paper $\langle .\rangle$ and $\bar{.}$ denote respectively vertical and horizontal averaging over the entire computational domain.

These characteristic times are also reported in table \ref{tab:siminfo} for all the DNS cases studied herein. For all the simulations we choose $Re_0=6000$ and $Pr=8$. As demonstrated in \cite{SCP15}, the character of the `turbulence' is qualitatively different at smaller $Re_0$. Also $Pr=8$ is approximately characteristic of thermally-stratified water, and $R \geq \sqrt{8}$ is sufficiently large that such flows are generically unstable to HWI \citep{Smyth_1988_Holmboe}.

\begin{table}
  \begin{center}
  \def~{\hphantom{0}}
  \begin{tabular}{rcccccccccccc}
     ~~Sim.~~ & $Ri_b$ & $R$ & $Ri_g(0)$	& $\sigma_r$ & $\lambda$ &
     $t_{2d}$ & $t_{3d}$ & $t_{rl}$ & $N_x$ & $N_y$ &  $N_z$ &  $N^c_z$	\\[3pt]
     ~~KHI~~  & 0.16  & 1    & 0.16 & 0.078 & 14.27 & 92  & 128 & 206 & 127  & 25 & 116  &  ~88 \\     
     R3-J016  & 0.16  & 2.83 & 0.45 & 0.027 & 16.15 & 164 & 222 & 318 & 145  & 26 & 116  &  ~88 \\     
     R5-J016  & 0.16  & 5    & 0.8  & 0.078 & ~9.67 & ~70 & 130 & 238 & ~87  & 26 & 116  &  ~88 \\     
     R10-J016 & 0.16  & 10   & 1.6  & 0.102 & ~7.85 & ~58 & 110 & 210 & ~71  & 26 & 116  &  ~88 \\
     R25-J016 & 0.16  & 25   & 4    & 0.104 & ~7.76 & ~76 & 124 & 212 & 107  & 39 & 164  &  132 \\
     R5-J008  & 0.08  & 5    & 0.4  & 0.074 & 11.64 & ~74 & 104 & 418 & 105  & 26 & 116  &  ~88 \\     
     R10-J008 & 0.08  & 10   & 0.8  & 0.103 & ~4.69 & ~56 & 110 & 364 & ~64  & 26 & 116  &  ~88 \\
     R5-J032  & 0.32  & 5    & 1.6  & 0.071 & ~7.14 & ~72 & 134 & 166 & ~85  & 26 & 116  &  ~88 \\     
     R10-J032 & 0.32  & 10   & 3.2  & 0.083 & ~5.93 & ~64 & 130 & 172 & ~54  & 26 & 116  &  ~88
  \end{tabular} 
  \caption{Details of the three-dimensional direct numerical simulations in which the total grid points are approximately $p^3N_xN_yN_z$ where $p=10$ is the order of Lagrange polynomial interpolants and $N_x$, $N_y$ and $N_z$ denote the number of spectral elements within the horizontal ($L_x$), spanwise ($L_y$) and vertical ($L_z$) extents of the computational domain. $N^c_z$ represents the number of elements within a central region of the domain with height $L^c_z=10$. Outside of $L^c_z$, the adjacent elements of the grid are gradually stretched by a factor of 1.25\%. In all these simulations, the initial Reynolds number $Re_0 = U_0 d/\nu = 6000$ and $Pr=\nu/\kappa = 8$, $L_x=\lambda $, $L_y=3$ and $L_z=30$.  $\sigma$ is the real part of the growth rate of the primary instability with a wavelength $\lambda$. The times $t_{2d}$, $t_{3d}$ and $t_{rl}$ have been defined in the text.}
  \label{tab:siminfo}
  \end{center}
\end{table}

We may also define two non-dimensionalized integral length scales associated with the time evolving shear and density layers, $\ell_u$ and $\ell_\rho$, as \citep{SCP15},
\begin{equation}
 \ell_u (t)     = \int_{z} \left(1-\overline{u}^2\right) dz,
 \qquad
 \ell_{\rho}(t) = \int_{z} \left[1-(\overline{\rho}-1)^2\right] dz,
 \label{eq:Iu_Irho}
\end{equation}
where  $\overline{u}$ and $\overline{\rho}$ represent non-dimensional, time-dependent velocity and density profiles of the background mean flow. Notice that at $t=0$, $\ell_u(0)/\ell_\rho(0) = d/\delta = R$ because $\overline{u}(z,0)$ and $\overline{\rho}(z,0)$ are initially defined in terms of hyperbolic tangents, as in \eqref{eq:tanhprof}. 
Now, $Re_b$ (as defined in \eqref{eq:Reb}) can also be interpreted as a ratio of two key physical length scales: the above-mentioned Ozmidov scale $\ell_O$, the largest vertical turbulent scale essentially unaffected by stratification, and $\ell_K$, the classical Kolmogorov length scale as
\begin{equation}
Re_b =
\left ( \frac{\ell_O}{\ell_K} \right )^{4/3} \quad
\ \ell_O= \alpha \left ( \frac{ \langle \bar{\epsilon'} \rangle}{[\langle N^2 \rangle ]^{3/2} } \right )^{1/2} \quad
\ \ell_K= \alpha \left ( \frac{\nu^3}{\langle \bar{\epsilon'} \rangle} \right )^{1/4},
\label{eq:lodef}
\end{equation}
where the scale factor $\alpha = (\ell_u/L_z)^{1/4}$ has been introduced to remove the scale dependence of the vertical averaging on computational domain size. As discussed
in more detail in \cite{SCP15}, both the regions where
the dissipation rate and the buoyancy frequency are substantially different from zero are typically concentrated
in $[-\ell_u/2, \ell_u/2]$, whereas the volume average is over the whole vertical domain $[-L_z/2, L_z/2]$, which 
extent is chosen precisely so that the far field regions are quiescent with no boundary effects.

In general, the mean flow may be distinguished from the perturbation component by employing (as mentioned above) horizontal averaging indicated by an overbar, which yields
\begin{eqnarray}
  \overline{u}(z,t)	 	= \overline{\boldsymbol{u}(x,y,z,t)}, \qquad
  \boldsymbol{u}'(x,y,z,t)	&=& \boldsymbol{u}(x,y,z,t) - \overline{u}(z,t)
 \label{eq:u=U+u'}
\\
  \overline{\rho}(z,t) = \overline{\rho(x,y,z,t)}, \qquad
  \rho' (x,y,z,t) &=& \rho(x,y,z,t) - \overline{\rho} (z,t) . \label{eq:rhomean}
\end{eqnarray}
In the remainder of this section, we will drop the explicit representation of temporal and spatial dependence for brevity.

An equilibrium state of stratified turbulence may be achieved when $dE_{ST}/dt \sim 0$ where $E_{ST}$ is the total energy budget of the stratified turbulence. Unlike homogeneous unstratified turbulence, $E_{ST}$ comprises both the turbulent kinetic energy $\mathcal{K}'$ and the available potential energy, $\mathcal{P}_A$, i.e.
\begin{equation}
 E_{ST} = \mathcal{K}' + \mathcal{P}_A.
 \label{eq:E_ST}
\end{equation}
The (volume-averaged) perturbation kinetic energy (associated with both two- and three-dimensional perturbations) may be defined as,
\begin{equation}
	\mathcal{K}'  =	0.5 \langle \overline{ \boldsymbol{u}' \cdot \boldsymbol{u}'} \rangle.
	\label{eq:KE'}
\end{equation}

The (volume-averaged) available potential energy, $\mathcal{P}_A$ may be identified as the difference between the total potential energy $\mathcal{P}$, and a reference background potential energy, $\mathcal{P}_B$, associated with a notional state that is statically stable and obtained by a continuous adiabatic rearrangement of the instantaneous density field, $\rho_*(z)$ \citep{Winters_1995, CP00}. $\mathcal{P}_A$ is available to be converted back into either $\mathcal{K}'$ or $\mathcal{P}_B$. These various (volume-averaged) components are defined by
\begin{gather}
 \mathcal{P}_A = \mathcal{P} - \mathcal{P}_B = Ri_b \langle (\overline{\rho} - \rho_*) z\rangle.
 \label{eq:APE}
\end{gather}
The required `sorting' procedure to find $\rho_*(z)$ has been implemented in parallel as described in \cite{SPM15} for application on distributed memory high performance computing clusters. All the simulations to be reported herein have been performed on a Blue Gene/Q system.

To investigate $dE_{ST}/dt$ we require the evolution equations for both $\mathcal{K}'$ and $\mathcal{P}_A$. For our closed, horizontally periodic system, these equations may be written as \citep{SP15},
\begin{gather}
 {d\over dt} E_{ST}	=  \mathbb{P} - \mathcal{M} - \mathcal{D},
 \label{eq:E_ST_evolution}		
 \\
  {d\over dt} \mathcal{K}'	= -\mathbb{B} + \mathbb{P} - \mathcal{D},
 \qquad
 {d\over dt} \mathcal{P}_A	= \mathbb{B} - \mathcal{M},
 \label{eq:TKE_APE}
\end{gather}
where the shear production due to the interaction of mean shear with Reynolds stresses, $\mathbb{P}$, the buoyancy flux, $\mathbb{B}$, the turbulent viscous dissipation, $\mathcal{D}$ and irreversible mixing, $\mathcal{M}$, are defined as \citep{SP15}
\begin{equation}
 \mathbb{P} = -\Big\langle {\partial\overline{u} \over \partial z} \hspace*{5pt} \overline{u'w'} \Big \rangle,
 \qquad
 \mathbb{B} = \frac{g}{\rho_r} \langle\, \overline{ \rho' w' } \, \rangle,
 \qquad
 \mathcal{D} = \langle\, \overline{\epsilon'} \,\rangle = 2\nu \langle \, \overline{  s'_{ij}s'_{ij} } \,\rangle ,
 \qquad
 \mathcal{M} = {d \mathcal{P}_B \over dt}-D_p,
 \label{eq:energy_components}
\end{equation}
in which $\boldsymbol{u}' = (u',v',w')$. Also $D_p=\kappa \langle N^2 \rangle$ represents the diffusive rate of increase in $\mathcal{P}_B$ in the absence of macroscopic motions.

Based on \eqref{eq:E_ST_evolution}, we may now define a quantitative measure for the approach of stratified turbulence towards equilibrium, i.e. when $dE_{ST}/dt \sim 0$ by defining the parameter $\mathscr{F}$ as
\begin{equation}
 \mathscr{F} = \frac{\mathbb{P} - \mathcal{M}}{\mathcal{D}}.
 \label{eq:F_ratio}
\end{equation}
See also \cite{Holt_1992} and \cite{Strang_2001} for similar definitions. In terms of this parameter, turbulence approaches equilibrium when $\mathscr{F} \sim 1$, while the turbulence may be considered to be growing or decaying when $\mathscr{F}>1$ or $\mathscr{F}<1$ respectively. The numerator of this parameter captures the residual turbulent kinetic energy production after irreversible conversion into potential energy through mixing, while the denominator is the viscous dissipation rate. Therefore, when this quantity is close to one, the flow is in equilibrium with viscous dissipation doing `just enough' to ensure $E_{ST}$ neither grows nor decays.

\section{Results}
\label{sec:results}

After a visual characterization of the DNS results in what follows, we will further analyze the induced turbulent flow in order to characterize the emergence of SOC and to discuss the implications of SOC for turbulence energetics and mixing within stratified shear flows.

\subsection{Evolution of HWI} 
\label{sec:evolution}

\begin{figure}
  \centerline{\includegraphics[width=\textwidth]{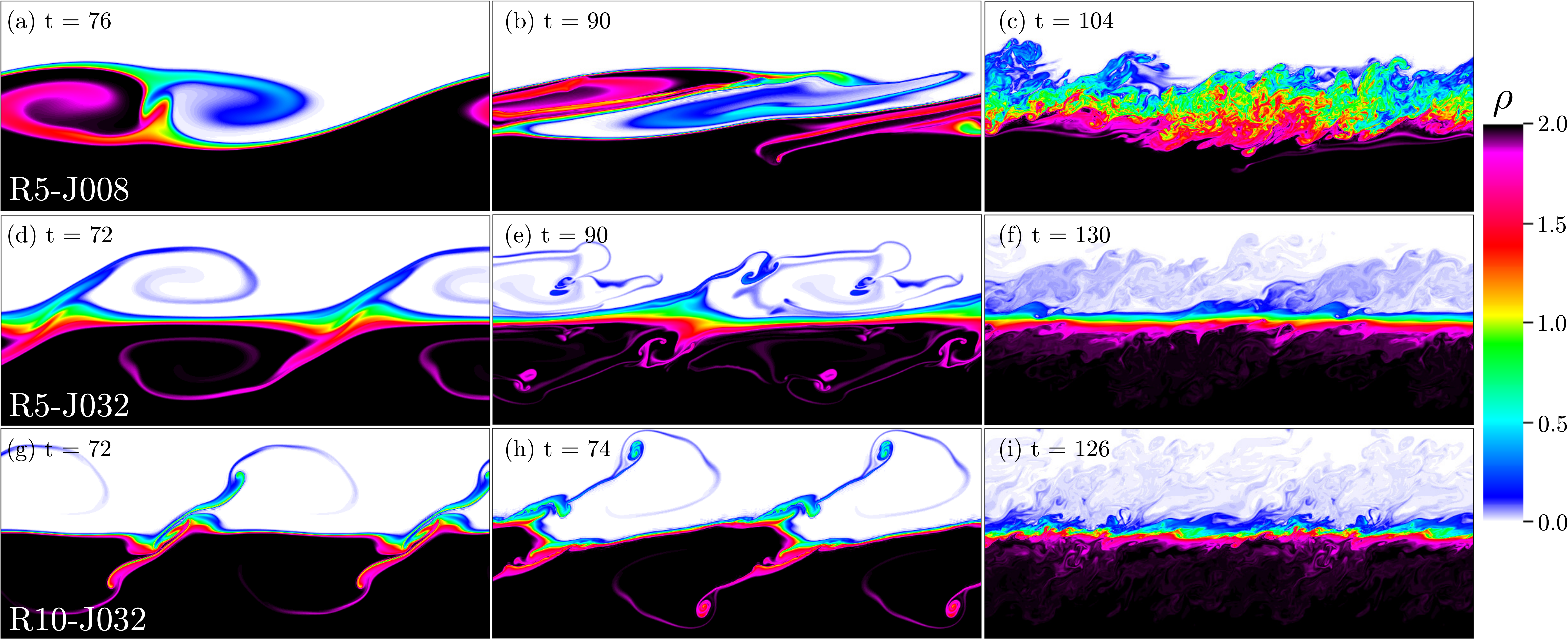}}
  \caption{Contour plots of density evolution at the indicated non-dimensional times for simulation R5-J008 (top), R5-J032 (middle) and R10-J032 (bottom) on the $x-z$ plane at the spanwise midpoint of the computational domain. The vertical extent in each panel is limited to $-2.5d\leq z\leq 2.5d$ ($L_z=30d$). The horizontal extents in (a,b,c) illustrate their corresponding $L_x$ while $x$-periodicity is invoked in other panels to result in identical horizontal extent (and hence aspect ratio) for all the panels in this figure. Refer to table \ref{tab:siminfo} for the characteristic times associated with each simulation.
}
  \label{fig:density}
\end{figure}

The growth and collapse of HWI for three of the simulations listed in table \ref{tab:siminfo} are illustrated in figure \ref{fig:density}. Cross-sections of the density distribution at the spanwise mid-plane of these three-dimensional flows are plotted at three different times. It is evident from panels (a,d,g) (near $t=t_{2d}$, see table \ref{tab:siminfo}), that increasing both $Ri_b$ and $R$ have noticeable and complicated effects on the form and shape of the saturated primary HWI. The observed `braid' and `billow' structure in figure \ref{fig:density}(a) is somewhat reminiscent of KHI. Therefore, perhaps unsurprisingly, it appears that HWI at $Ri_b=0.08$ is close to the transition between KHI and HWI (see \cite{Smyth_1989,Hogg_2003} for further discussion) while at $Ri_b=0.32$ the interface remains close to horizontal (in particular there is no `braid' between adjacent `billows') with the interface being perturbed by  `wisps', ejected from interfacial cusp-shaped counter-propagating waves, apparently induced by vortices above and below the interface. Increasing $R$ through sharpening the density interface (equivalent to reducing $\ell_\rho(0)$ in equation \eqref{eq:Iu_Irho}) appears to enhance the characteristic `wisping' further, with more complex near-interface structure developing. Indeed, these stronger `wisps' tend to destabilize the flow further such that the induced turbulence after the full break down of the primary instabilities is more vigorous at higher $R$ (c.f. panels (f,i)). On the other hand, if the bulk stratification is increased (higher $Ri_b$) for a given initial thickness ratio $R$,  the ensuing turbulence is less vigorous (c.f. panels (c,f)). 


\cite{Smyth_etal_2007}, who have also investigated the effect of varying $Ri_b$ (at a fixed $R$) observed that at $Ri_b=1/3$ (and $R=3$) the flow `never becomes turbulent'. The difference between their results and ours (as shown for example in figure \ref{fig:density}f) may be explained by noting that their initial Reynolds number is approximately twenty times smaller than $Re=6000$ and, as discussed in \cite{SCP15}, increasing $Re$ has profound impacts not only on the transition to turbulence itself but also on the mixing and spectral properties of the ensuing turbulent flow. Therefore, based on the discussions of \cite{SCP15}, we expect that if $Re$ were further increased beyond $Re=6000$, the flow would become increasingly more energetically turbulent because the growth rate of three-dimensional secondary instabilities would be expected to approach their asymptotic `inviscid' rates.

\subsection{HWI-induced turbulence at quasi-equilibrium} 
\label{sec:equilibrium} 

The slowly-evolving and long-lived nature of the turbulence engendered by HWI, (as discussed on the basis of figure \ref{fig:longevity}) might suggest the existence of a \emph{`quasi-equilibrium'} state. It is tempting to argue that this state corroborates `a kind of equilibrium' as postulated by \cite{Turner_1979} for `strongly stratified' flows. It must be mentioned, however, that the flows studied in this paper are examples of freely evolving dissipative systems, unlike a gravity current on a relatively steep slope (see figure (4.19) of \citep{Turner_1979}), or the equatorial undercurrents \citep{Smyth_etal_2013_marginal} which are examples of forced-dissipative systems. A statistically stationary equilibrium state with $\mathscr{F}(t)\sim 1$ (see equation \eqref{eq:F_ratio}) would only be expected in such forced-dissipative circumstances. 

\begin{figure}
  \centering 
  \includegraphics[width=\textwidth]{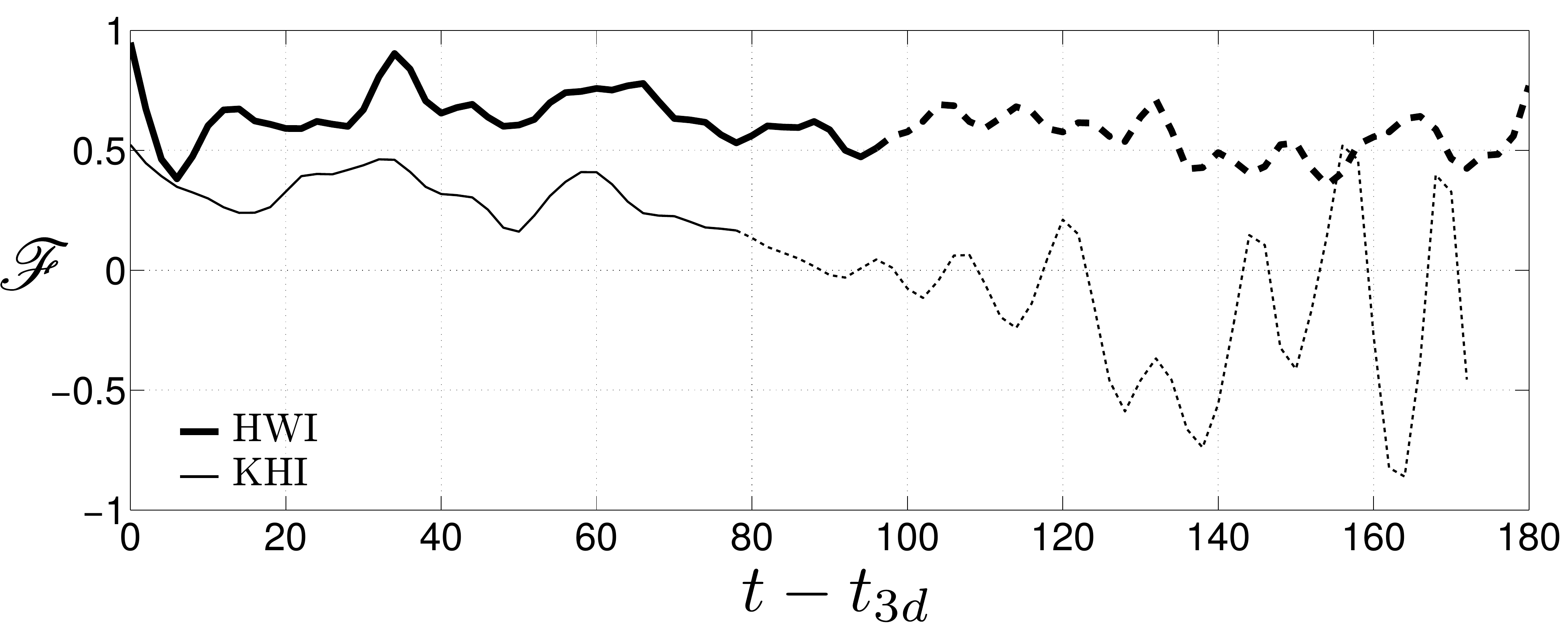}
  \caption{The variation with time of $\mathscr{F}$ after the onset of fully three-dimensional flow (i.e. $t\geq t_{3d}$). The re-laminarized phase beginning at $t_{rl}$ is indicated by dashed lines.}
    \label{fig:F}%
\end{figure}

Figure \ref{fig:F} illustrates the time variations of $\mathscr{F}$ for $t \geq t_{3d}$ associated with both HWI (case R3-J016) and KHI. The fully turbulent and re-laminarized stages are indicated respectively by solid and dashed curves.  As discussed in \cite{SCP15}, the HWI-induced turbulence achieves its longevity by the localization of mixing due to scouring motions on the flanks of the density interface. In other words, HWI `self-organises' the location of scouring motions in such a way as to survive for as long as possible by remaining in a state of quasi-equilibrium (that is perhaps close to marginal instability, in some sense) with relatively constant $\mathscr{F}$. On the other hand, the KHI involves a relatively more sudden mixing event that is localized around the interface by overturning motions leading to a decreasing trend for $\mathscr{F}$. As a result, turbulence is relatively short-lived in the latter case.

The quasi-equilibrium behavior of HWI-induced turbulence appears to be a robust characteristic property of stratified sheared turbulence arising from such relatively sharp density interfaces embedded in relatively deep shear layers, since all our HWI cases (regardless of their initial conditions, such as the initial values chosen for $Ri_b$ or $R$, provided of course that the flow is susceptible to a primary HWI) deliver relatively constant values of $\mathscr{F}$. This behavior is categorically different from that characteristic of KHI. (The variation of $\mathscr{F}(t)$ for all HWI cases is not presented here for brevity.) In summary, it appears that quasi-equilibrium conditions are generically achieved and sustained by the collapse of HWI into turbulence, even for arbitrarily large values of stratification.

\subsection{HWI-induced turbulence and self-organised criticality} 
\label{sec:soc} 

Here we investigate the mean and turbulent characteristics of the flow induced by HWI in order to describe the basis on which these characteristics might be understood in terms of the SOC ansatz.

\subsubsection{Self-organization towards a critical gradient Richardson number} 
\label{sec:Rig} 

\begin{figure}
    \centerline{\includegraphics[width=0.85\textwidth]{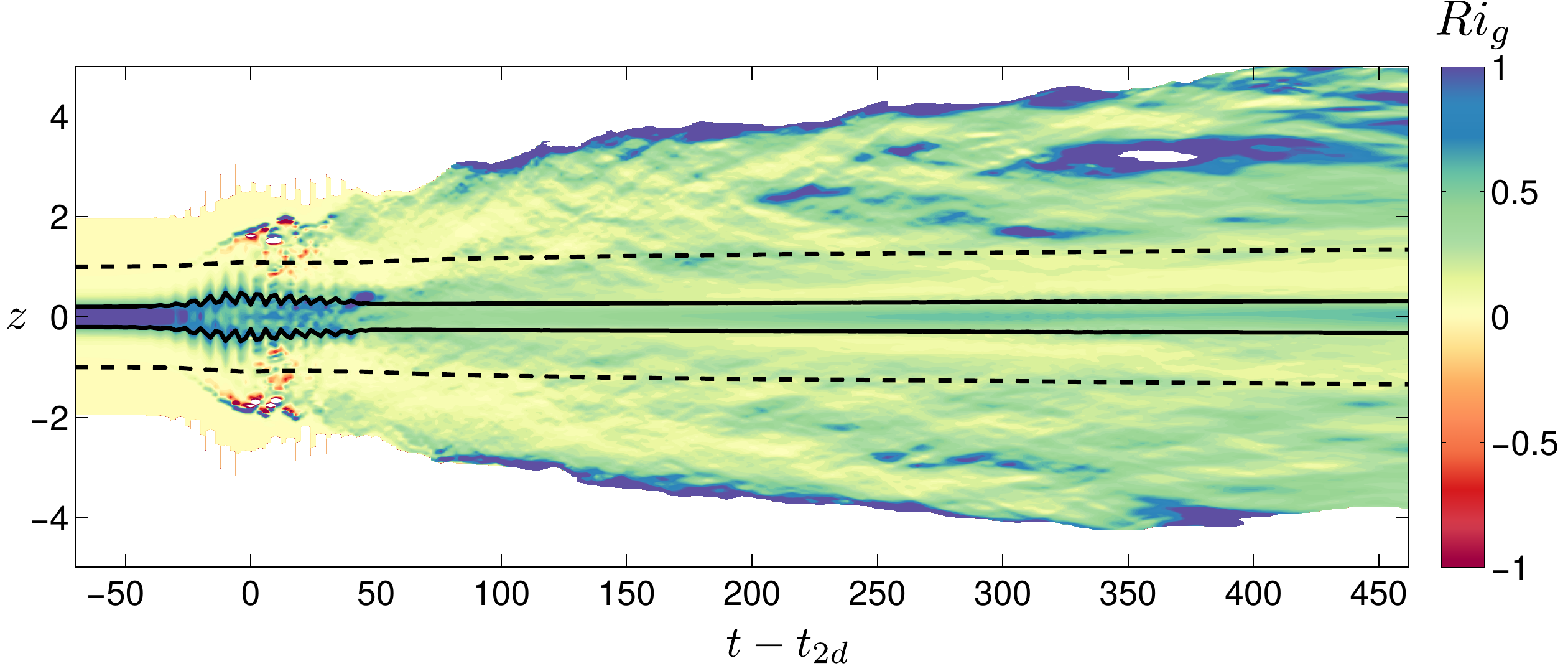}}
    \caption{The variation with time $t$ and vertical coordinate $z$ of $Ri_g(z,t)$ for case R5-J032. For reference, the upper and lower extents of the integral length scales $\ell_u$ (dashed lines) and $\ell_\rho$ (solid lines) are also plotted.}
    \label{fig:Rig_zt}%
\end{figure}

\begin{figure}
    \centerline{\includegraphics[width=\textwidth]{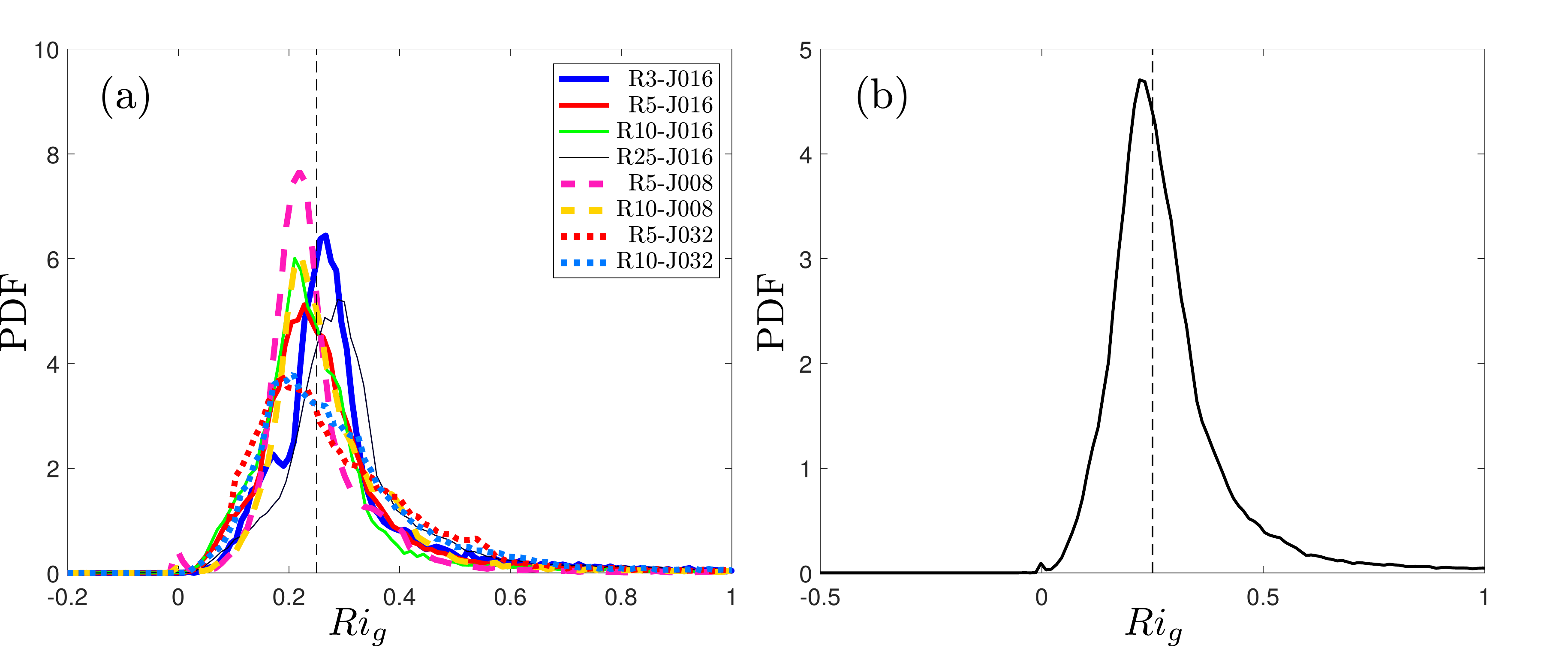}}
    \caption{(a) The probability density function of $Ri_g(z,t)$ for $t\geq t_{3d}$ for each HWI case, plotted with different line types as listed in the legend. (b) The probability density function of $Ri_g(z,t)$ for $t\geq t_{3d}$, aggregated from all HWI cases combined.}
    \label{fig:Rig_PDF}%
\end{figure}

First, we analyze the local (in space and time) values of the gradient Richardson number, $Ri_g(z,t)$, as defined in \eqref{eq:Rig}. Figure \ref{fig:Rig_zt} illustrates the temporal evolution of the vertical profile of $Ri_g(z,t)$ for the case R5-J032. For reference, this figure also illustrates the upper and lower extents of the integral length scales $\ell_u$ (dashed) and $\ell_\rho$ (solid). The probability density functions (PDF) of such local values of $Ri_g(z,t)$, combined for all times $t\geq t_{3d}$ and for all cases susceptible to HWI are plotted in figure \ref{fig:Rig_PDF}, presented both individually for each case (in panel (a)) and aggregated together for all cases (in panel (b)). In both these figures, 
spurious large values of $Ri_g$ (as a result of a close to undetermined $0/0$ condition, i.e. if $Ri_g>100$) as well as zero $Ri_g$ values associated with unstratified, non-turbulent regions (with negligible turbulent dissipation rates $\overline{\epsilon'(z)} < Pr D_p$ where $D_p$ is the molecular dissipation rate) have been excluded from our analysis, thus virtually completely eliminating a trivial peak in the PDF at $Ri_g=0$, associated with unstratified, non-turbulent regions.

As shown in figure \ref{fig:Rig_zt}, $Ri_g(z,t)$ has a complex spatiotemporal structure, and hence it is unclear whether the interfacial value of $Ri_g$ (i.e. $Ri_g(0,t)$), that initially represents the maximum (for all $z$) of $Ri_g(z,t=0)$ for flows susceptible to HWI, is an adequate diagnostic parameter to characterize the induced turbulent mixing. In particular, classifying the stratification as being `strong' or not based only on this specific value seems potentially misleading. However, the PDF of $Ri_g(z,t)$ (for all $z$, and for sufficiently large $t \geq t_{3d}$ so that the flow may be characterized as being turbulent) in figure \ref{fig:Rig_PDF} demonstrates a striking characteristic distribution in which the vast majority of local $Ri_g$ values lie in the proximity of $Ri_g \sim 0.2-0.25$, implying that a large proportion of these turbulent flows share similar mean flow characteristics. Indeed, this observed characteristic feature appears to be a robust property of HWI-induced turbulence \emph{irrespective} of the initial density and shear layer depths or the bulk stratification and is at least somewhat reminiscent of oceanic observations in the Pacific equatorial undercurrents (see, for example, figure 2 of \cite{Smyth_etal_2013_marginal}). 

\begin{figure}
  \centering 
  \includegraphics[width=\textwidth]{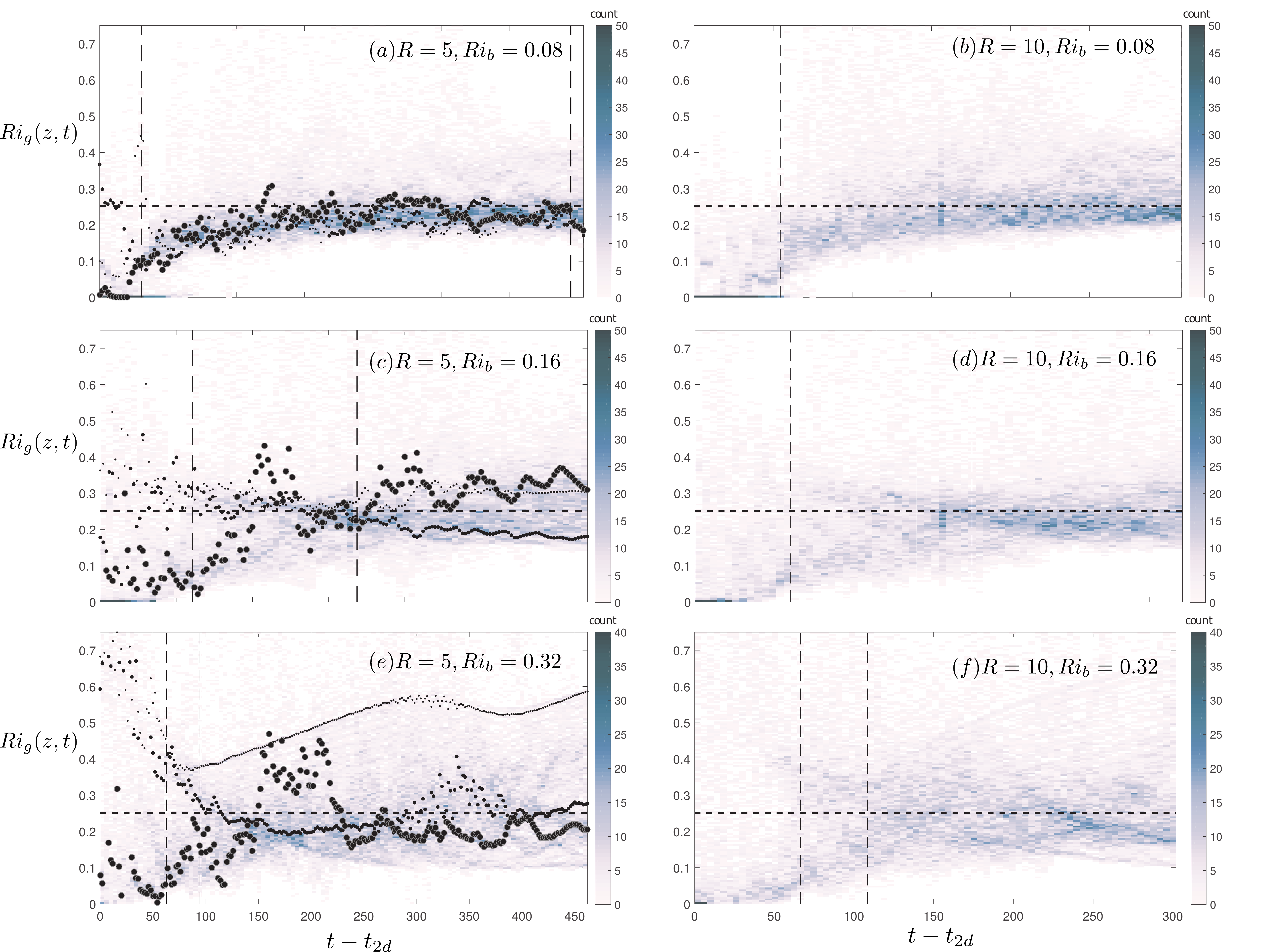}
  \caption{Two-dimensional histograms of $Ri_g(z,t)$ as a function of time for $t\geq t_{2d}$, for six DNS cases for flows susceptible to HWI (as listed in table \ref{tab:siminfo}) with $R=5$ in the left column (a,c,e) and $R=10$ in the right column (b,d,f); $Ri_b=0.08$ in the top row (a,b), $Ri_b=0.16$ in the middle row (c,d) and $Ri_b=0.32$ in the bottom row (e,f). In the left column, the variations with time of $Ri_g(z=0,t)$, $Ri_g(z=\ell_\rho/2,t)$ and $Ri_g(z=\ell_u/2,t)$ are also plotted by increasingly larger sizes of `$\bullet$'. The first vertical dashed line in each panel marks the time $t_{3d}$. The second vertical dashed line (if present) marks the time $t_{rl}$. The horizontal dashed line denotes $Ri_g=0.25$. }
    \label{fig:SOC}%
\end{figure}

This characteristic distribution of $Ri_g$ for flows susceptible to primary HWI leads us to conjecture that $Ri_g \simeq 1/4$ is a critical state that acts as an `attractor' towards which the flow tends to self-organize without any external tuning. To test  this conjecture further, in figure \ref{fig:SOC} we plot the time-dependence of the two-dimensional histograms associated with the occurences of $Ri_g$, for times  $ t \geq t_{2d}$. Note that the $Ri_g$ bins used to construct the PDFs shown in figure \ref{fig:Rig_PDF} aggregate data for all times  $t\geq t_{3d}$ leading to a one-dimensional histogram underlying the (normalized) PDF of $Ri_g$, whereas figure \ref{fig:SOC} includes two-dimensional histograms in which both $Ri_g$ and time (subsequent to $t_{2d}$) are appropriately binned. For reference, the panels in the left column of  this figure also plot $Ri_g(z=0,t)$, $Ri_g(z=\ell_\rho/2,t)$ and $Ri_g(z=\ell_u/2,t)$ marked by `$\bullet$' with respectively increasing symbol sizes (i.e. locations farther away from the interface correspond to larger symbols), which highlight the corresponding values of $Ri_g$ at these characteristic vertical locations.

Figure \ref{fig:SOC} shows the eventual concentration of $Ri_g$ near to the value of $1/4$ ($Ri_g=1/4$ is indicated by a horizontal dashed line), either by a gradual increase of the $Ri_g$ distribution from lower values  for flows with smaller values of $Ri_b$ (i.e. figure \ref{fig:SOC}(a,b)) or by merging of a bimodal distribution with peaks above and below 1/4 into a unimodal distribution with a peak near 1/4 for flows with larger values of $Ri_b$ (most apparent in figures \ref{fig:SOC}(d,f)). Essentially, the flow \emph{self-organizes} to reach the critical state associated with `marginal instability' with $Ri_g\sim 1/4$.

Again, recall that for our chosen initial velocity and density hyperbolic tangent profiles (\ref{eq:tanhprof}) with sufficiently large $R=\sqrt{8}$ such that the flow is susceptible to primary HWI, the initial profile of $Ri_g$ is maximum at the interface (i.e. $z=0$) and decreases towards the `flanks' of the density and shear layer (i.e. $z=\pm \ell_\rho/2$ and $z=\pm \ell_u/2$). Hence, for flows susceptible to primary HWI, $Ri_g(\pm \ell_u/2,0) \ll Ri_g(\pm \ell_\rho/2,0) < Ri_g(0)$, and the difference between these different values becomes larger as $Ri_b$ or $R$ increases. As shown in figure \ref{fig:SOC}(a,c,e), by the time $t=t_{2d}$, $Ri_g(0,t_{2d})$ has decreased from its initial value due to the pre-turbulent mixing that reduces the density gradient at the interface (see also figure \ref{fig:density}). This is especially noticeable in the relatively weakly stratified DNS cases with $Ri_b=0.08$ in which $Ri_g(0,t_{2d}) \ll 1/4$ (see \eg figure \ref{fig:SOC}(a)) due to the nontrivial perturbation  of the density interface, as shown in  figure \ref{fig:density}(b). As the flow evolves towards $t=t_{3d}$ (marked by the first vertical dashed line in figure \ref{fig:SOC}), this interfacial value of $Ri_g$ also evolves towards 1/4 either from below (figure \ref{fig:SOC}(a,b) at $Ri_b=0.08$) or from above (figure \ref{fig:SOC}(c,d) at $Ri_b=0.16$). If the turbulence is sufficiently strong (as it is for the simulations with $Ri_b=0.08$ and $Ri_b=0.16$), the interfacial value of $Ri_g$ as well as those within the shear layer (i.e. for $|z| \leq  \ell_u/2$) and the density interface (i.e. for $|z| \leq \ell_\rho/2$) are attracted towards the critical state with $Ri_g=1/4$.

However, at $Ri_b=0.32$ (see \eg case R5-J032 as plotted in figure \ref{fig:SOC}(e)), the overall turbulence intensity is not as strong and hence turbulent mixing remains relatively more localized at the flanks of the shear layer than at the density interface, largely because turbulent motions are not able to penetrate sufficiently close to that density interface. This is the physical reason why  $Ri_g(0,t)$ increases again during the re-laminarization stage, as shown in figure \ref{fig:SOC}(e).  (The onset of re-laminarization, at $t_{rl}$, is marked, when appropriate, by the second vertical dashed line in figure \ref{fig:SOC}). Nonetheless, even for the flow with  $Ri_b=0.32$,  the flow self-organizes, as in the other cases in figure \ref{fig:SOC}, such that a sharply peaked distribution of $Ri_g$ is achieved in the near vicinity of $Ri_g\sim 1/4$, consistently with all the other cases  that there is an attractor associated with a critical marginal state with $Ri_g \simeq 1/4$.

\begin{figure}
  \centering 
  \includegraphics[width=\textwidth]{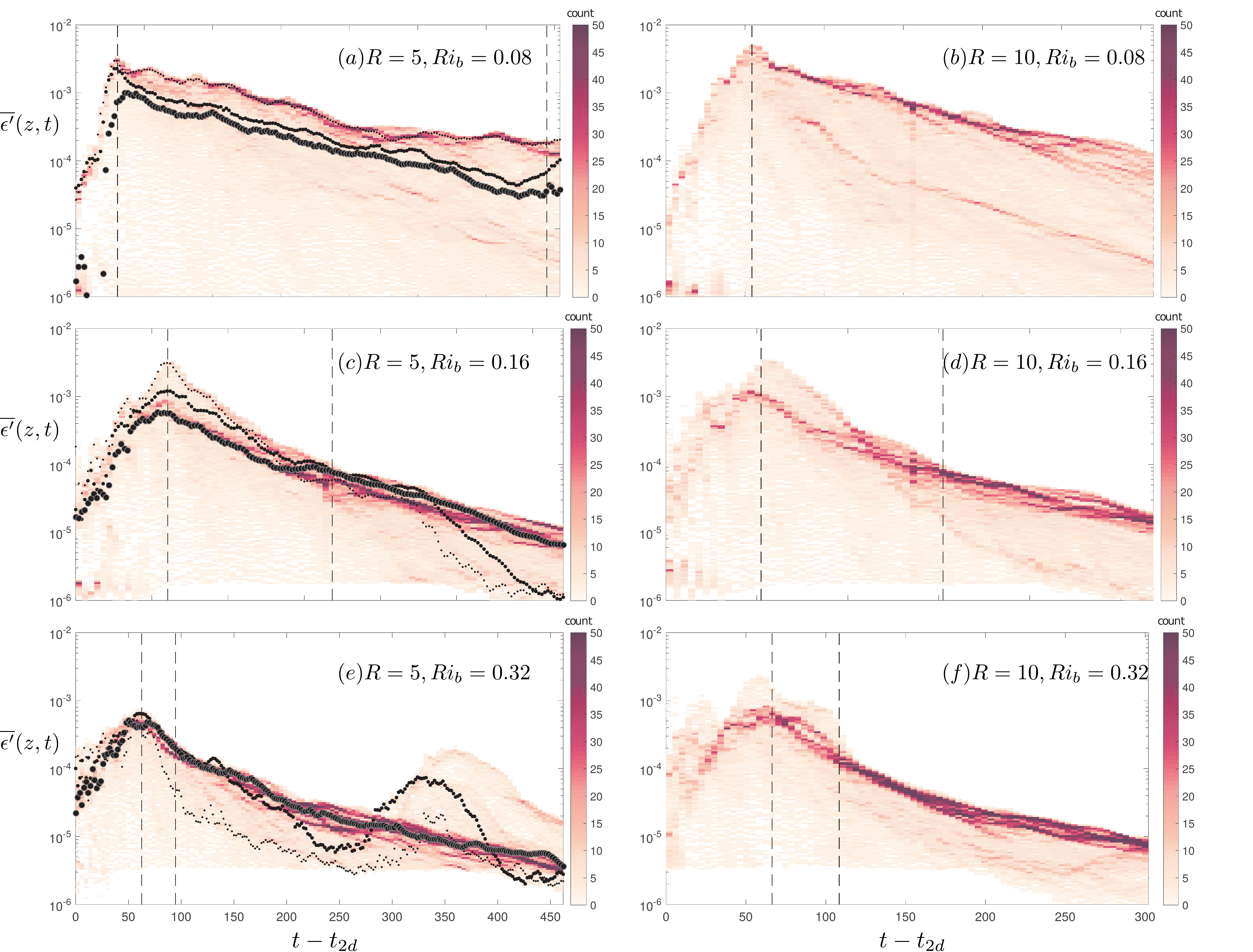}
  \caption{Two-dimensional histograms of horizontally averaged turbulent dissipation $\overline{\epsilon'}(z,t)$ (as defined in \eqref{eq:epsilon_z}) as a function of time for $t\geq t_{2d}$, for six DNS cases for flows susceptible to HWI (as listed in table \ref{tab:siminfo}) with $R=5$ in the left column (a,c,e) and $R=10$ in the right column (b,d,f); with $Ri_b=0.08$ in the top row (a,b), $Ri_b=0.16$ in the middle row (c,d) and $Ri_b=0.32$ in the bottom row (e,f). In the left column, the variations with time of $\overline{\epsilon'}(z=0,t)$, $\overline{\epsilon'}(z=\ell_\rho/2,t)$ and $\overline{\epsilon'}(z=\ell_u/2,t)$ are also plotted by  increasingly larger sizes of `$\bullet$'. The first vertical dashed line in each panel marks the time $t_{3d}$. The second vertical dashed line (if present) marks the time $t_{rl}$.}
    \label{fig:SOC_eps}%
\end{figure}

We further investigate the inferred self-organization mechanism by considering various quantities associated with the horizontally-averaged turbulent dissipation $\overline{\epsilon'}(z,t)$ (as defined in \eqref{eq:epsilon_z}). Similarly to figure \ref{fig:SOC}, in figure   \ref{fig:SOC_eps} we plot the time-dependence of the two-dimensional histograms associated with the occurence of binned values of $\overline{\epsilon'}(z,t)$. Once again, the corresponding values of $\overline{\epsilon'}(0,t)$, $\overline{\epsilon'}(\ell_\rho/2,t)$ and $\overline{\epsilon'}(\ell_u/2,t)$ are also overlaid in the panels in the left column. At $Ri_b=0.08$, the `avalanches'
(inevitably associated with locally enhanced dissipation) of the sandpile model of the SOC  process are mostly concentrated in the vicinity of the density interface, (whose equilibrium location is at $z=0$) where the turbulent dissipation is also (globally across the whole mixing layer) maximum.

Upon doubling the bulk stratification to $Ri_b=0.16$, the turbulent `avalanches' follow  a different path to regulate the horizontally-averaged flow towards $Ri_g \sim 1/4$. The avalanches  become mostly concentrated between $\ell_\rho<z<\ell_u$, although their associated dissipation is smaller than that at the interface. However, by the time the flow re-laminarizes and reaches the turbulent analogue of the critical `angle of repose' of the `sandpile' associated with $Ri_g=1/4$ (marked by the second vertical dashed line), dissipation becomes distributed differently such that $\overline{\epsilon'} (\ell_u/2,t) >   \overline{\epsilon'} (\ell_\rho/2,t) >  \overline{\epsilon'} (0,t)$ which is required to accommodate the maintenance of the critical state.

Further doubling the bulk stratification to $Ri_b = 0.32$ provides a slightly different path to self-regularization. Similar to the previous case, at this value of $Ri_b$ the peak value of the histogram of the horizontally averaged turbulent dissipation rate occurs between $\ell_\rho<z<\ell_u$. Nevertheless, and in contrast to the previous case with lower $Ri_b$, this location of peak probability also corresponds to the maximum turbulent dissipation. This self-regulated localization of turbulent dissipation appears to be a key factor in arriving at a mean flow that is characterized by a critical state with $Ri_g \sim 1/4$. The localization of dissipation and mixing at the upper and lower `flanks' of the interface is mediated by `scouring' motions, evident in figure \ref{fig:density} and  characteristic of HWI-induced turbulence as discussed in \cite{SCP15} and reviewed in \S\ref{sec:intro}.

The attractor associated with the marginal state of $Ri_g \sim 1/4$ appears to be a \emph{robust} characteristic of HWI-induced turbulence regardless of the initial bulk stratification. In fact, scouring motions evidently re-emerge in the case R5-J032 during $300<t-t_{2d}<350$ when a temporary rise in $\overline{\epsilon'}(\ell_\rho/2,t)$ (see figure \ref{fig:SOC_eps}(e)) occurs. Nonetheless, $Ri_g(\ell_\rho/2,t)$ becomes re-attracted towards the critical state with $Ri_g\sim1/4$ after a secondary transition period of self-regulation.  To use the sandpile analogy, further `sand' is added by depositing more kinetic energy into the flow, thereby causing deviations from the critical state and thus re-emergence of `avalanches' (here in the actual form of scouring mixing events) which relatively rapidly re-organizes the mean flow towards the critical state. Clearly, this process requires no external tuning for self-organization.

\subsubsection{Self-organization of energetics and mixing} 
\label{sec:eff} 
%
Considering the entire life-cycle of a flow susceptible to a primary instability (i.e. beginning and ending with a laminar state, denoted here as occurring at $t=0$ and $t=\tau$ respectively), there must be no net gain in the total energy of the stratified turbulence, defined in \eqref{eq:E_ST}, i.e. $\Delta E_{ST} = 0$. This also implies that there must be no net gain in the turbulent kinetic energy and available potential energy, defined in \eqref{eq:TKE_APE}, i.e. $\Delta \mathcal{K}'=0$, $\Delta \mathcal{P}_A = 0$. Equivalently, integrating \eqref{eq:E_ST_evolution} and \eqref{eq:TKE_APE} over the entire life-cycle yields
\begin{eqnarray}
  \widetilde{\mathbb{P}} &=& \widetilde{\mathcal{M}}  + \widetilde{\mathcal{D}},
  \label{eq:life_cycle_balance}
   \\
  \widetilde{\mathbb{P}} &=& \widetilde{\mathbb{B}}   + \widetilde{\mathcal{D}},
  \label{eq:life_cycle_balance_TKE}
  \\
  \widetilde{\mathbb{B}} &=& \widetilde{\mathcal{M}},
  \label{eq:life_cycle_balance_APE}
\end{eqnarray}
in which the \emph{life-cycle-averaged} quantities are denoted by a tilde, \eg $\widetilde{\mathcal{M}} = 1/\tau \int_0^\tau \mathcal{M}(t) \, dt $ indicates the life-cycle average of the irreversible mixing rate. More generally, we may employ the following definitions for cumulative time-dependent quantities,
\begin{equation}
 \mathcal{M}_c(t) = \frac{1}{t}\displaystyle\int_{0}^t \mathcal{M} \,dt, \quad
 \mathbb{B}_c(t)  = \frac{1}{t}\displaystyle\int_{0}^t \mathbb{B}  \,dt, \quad 
 \mathcal{D}_c(t) = \frac{1}{t}\displaystyle\int_{0}^t \mathcal{D} \,dt, \quad 
 \mathbb{P}_c(t)  = \frac{1}{t}\displaystyle\int_{0}^t \mathbb{P}  \,dt.
 \label{eq:cum_quantities}
\end{equation}
It is important to note that based on \eqref{eq:life_cycle_balance_APE}, only $\widetilde{\mathcal{M}} = \widetilde{\mathbb{B}}$ ($=\mathcal{M}_c(\tau) = \mathbb{B}_c(\tau)$) and in general $\mathcal{M}_c(t) \neq \mathbb{B}_c(t)$ as captured by the variability of $\mathcal{P}_A(t)$ due to the transient reversible processes associated with the buoyancy flux, $\mathbb{B}$.
%

Equation \eqref{eq:life_cycle_balance} implies that the total shear production of turbulent kinetic energy due to the interaction of perturbation Reynolds stresses with the mean flow ultimately either contributes to the cumulative irreversible mixing, or is lost due to turbulent viscous dissipation. Most critical in the above balance equations is the partitioning of these processes, since over the entire period both
\begin{equation}
  1 = \frac{\widetilde{\mathcal{M}}}{\widetilde{\mathbb{P}}} + \frac{\widetilde{\mathcal{D}}}{\widetilde{\mathbb{P}}}, 
  \label{eq:life_cycle_balance_fraction}
 \end{equation}
and
\begin{equation}
   1 =\frac{\widetilde{\mathbb{B}}}{\widetilde{\mathbb{P}}} + \frac{\widetilde{\mathcal{D}}}{\widetilde{\mathbb{P}}}, 
  \label{eq:life_cycle_balance_fraction2}
\end{equation}
%

\begin{figure}
  \centering 
  \includegraphics[width=\textwidth]{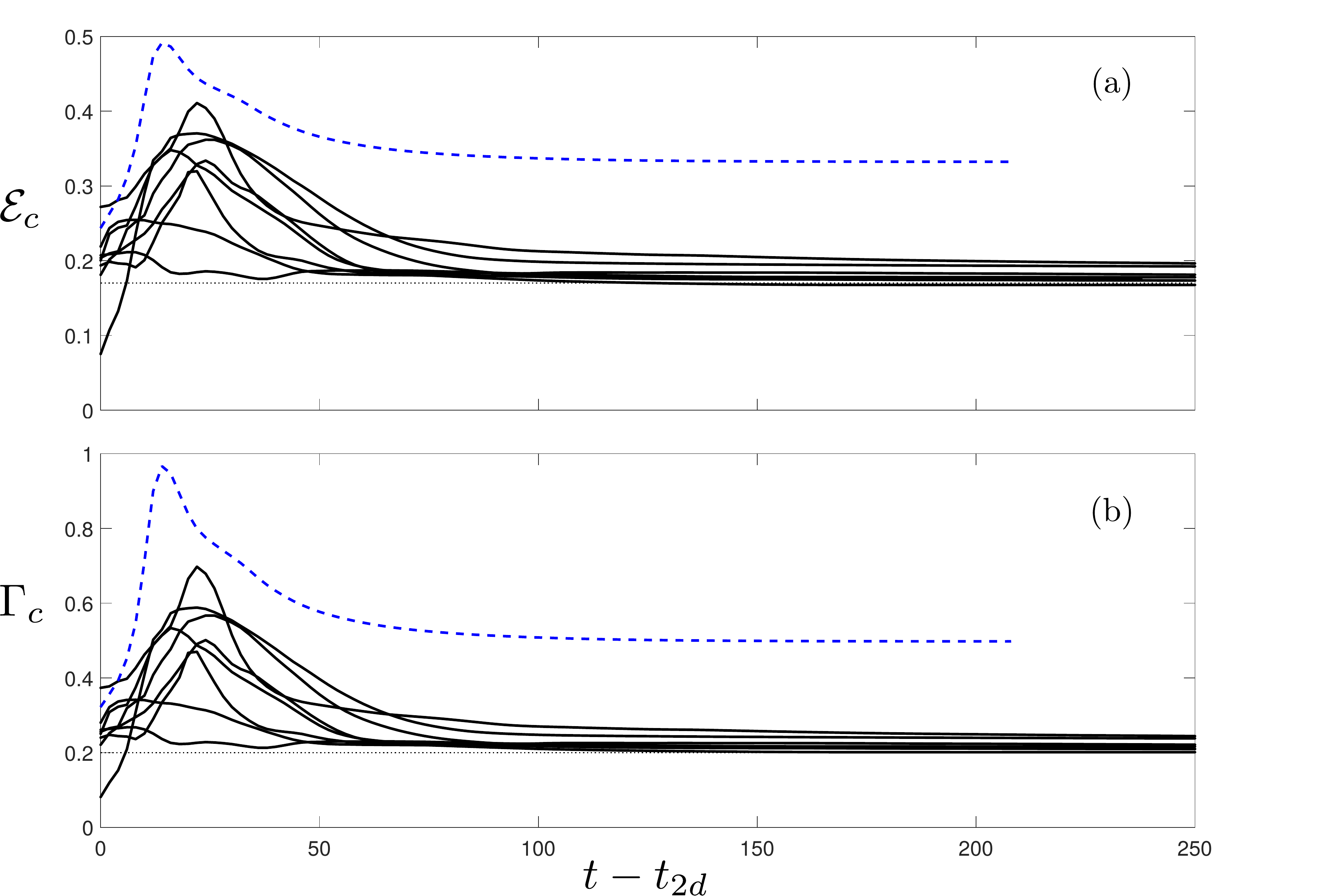}
  \caption{Variation with time of the cumulative quantities associated with the partitioning of stratified turbulence energy as defined in \eqref{eq:cum_quantities}: (a) cumulative mixing efficiency $\mathcal{E}_c(t)$, as defined in \eqref{eq:cumulative_efficiency}; and (b) cumulative turbulent flux coefficient $\Gamma_c(t)$, as defined in \eqref{eq:cumulative_turbulent_flux},  for the eight cases associated with HWI-induced turbulence (solid lines) and the case associated with KHI-induced turbulence (dashed line) as listed in table \ref{tab:siminfo}. Horizontal dashed lines correspond to the upper bound proposed by \cite{Osborn_1980} such that $\Gamma_c = 0.2$, and hence $\mathcal{E}_c = 1/6$.}
    \label{fig:cum_soc}%
\end{figure}

The first term on the right hand side of \eqref{eq:life_cycle_balance_fraction} is the life-cycle-averaged mixing efficiency, $\widetilde{\mathcal{E}}$, defined as
\begin{equation}
 \widetilde{\mathcal{E}} = \frac{\widetilde{\mathcal{M}}}{\widetilde{\mathcal{M}} + \widetilde{\mathcal{D}}} = \frac{\widetilde{\mathbb{B}}}{\widetilde{\mathbb{B}} + \widetilde{\mathcal{D}}} = \widetilde{Ri}_f,
 \label{eq:Rif-life-cycle-averaged}
\end{equation}
in which the life-cycle-averaged \emph{flux} Richardson number, $\widetilde{Ri}_f$ is the first term on the right hand side of \eqref{eq:life_cycle_balance_fraction2} and its equality to $\widetilde{\mathcal{E}}$ follows from \eqref{eq:life_cycle_balance_APE}.

In general, we can define a cumulative mixing efficiency $\mathcal{E}_c(t)$  for all $t \in [0,\tau]$ as 
\begin{equation}
 \mathcal{E}_c(t) = \frac{ \mathcal{M}_c(t)}{ \mathcal{M}_c(t)  + \mathcal{D}_c(t)},
 \label{eq:cumulative_efficiency}
\end{equation}
following \cite{CP00}. Obviously, $\mathcal{E}_c(\tau) = \widetilde{\mathcal{E}}$. One may also equivalently define $\Gamma_c(t)$ as the cumulative \emph{turbulent flux coefficient}, 
\begin{equation}
 \Gamma_c(t) = \frac{\mathcal{M}_c(t)}{\mathcal{D}_c(t)} = \frac{\mathcal{E}_c(t)}{1-\mathcal{E}_c(t)}.
 \label{eq:cumulative_turbulent_flux}
\end{equation}

Figure \ref{fig:cum_soc} plots the time evolution of $\mathcal{E}_c(t)$ and $\Gamma_c(t)$ for all the cases with HWI-induced turbulence (solid curves) and the single case with KHI-induced turbulence (dashed curve) as listed in table \ref{tab:siminfo}. For comparison, the canonical values of mixing efficiency and flux coefficient commonly employed by oceanographers and proposed as an upper bound by \cite{Osborn_1980}, i.e. $\mathcal{E}_c = 1/6$ or $\Gamma_c = 0.2$, are also shown by dashed lines in figures \ref{fig:cum_soc}(a,b).

It is quite  startling to observe that for all HWI-induced turbulence cases, irrespective of the initial conditions of scale ratio $R$ and bulk Richardson number $Ri_b$, the cumulative mixing efficiency approaches a life-cycle-averaged value of $\mathcal{E}_c(\tau) \sim 1/6$, implying  that the turbulent flux coefficient tends to the upper bound value $\Gamma_c (\tau) \sim 0.2$ proposed by \cite{Osborn_1980}. This generic behavior of HWI-induced turbulence must be contrasted with that of KHI-induced turbulence which for the case investigated here approaches the much larger limit $\mathcal{E}_c (\tau) \sim 0.35$  (and equivalently $\Gamma_c(\tau) \sim 0.5$), values that are known to be case-dependent and hence very sensitive to the specified initial conditions.

For example, this relatively high mixing efficiency associated with KHI is known to vary non-monotonically with the initial bulk Richardson number $Ri_b$ \citep{MCP13, SP15} and to decrease with the molecular Prandtl number $Pr$ \citep{SPM15}. Such sensitivity to the `external tuning parameters' follows from the inherently transient and highly delicate nature of the overturning of the primary Kelvin-Helmholtz billow, significantly modulated by the emergence of three-dimensional secondary instabilities, which are in turn highly dependent on the external parameters \citep{MP13, SPM15}. The overturning process is absent from the HWI-induced turbulence and we believe this fundamental difference explains many of its robustly \emph{universal} characteristics.

We find that, based on the SOC ansatz, the HWI-induced turbulence is self-organised such that an apparently universal energetics partitioning emerges, surprisingly consistent with the upper bound proposed by \cite{Osborn_1980}. In contrast with the more commonly considered KHI-induced turbulence, it appears that the underlying assumption of stationarity of Osborn is more closely approximated by the quasi-equilibrium of HWI-induced turbulence. For such flows,  regardless of the initial conditions (i.e. without external tuning), the energy of stratified turbulence, $E_{ST}$ is partitioned such that a critical cumulative turbulent flux coefficient of $\Gamma_c \sim 0.2$ is reached, consistently with the upper bound partitioning postulated by  \cite{Osborn_1980}.

It is very important to appreciate  that this value of $\Gamma_c$ (in this class of flows at least) is associated with the flow self-organizing towards a critical mean state with $Ri_g(z,t) \sim 1/4$ throughout much of the flow, whereas the arguments presented in \cite{Osborn_1980} relied at least partially on the semi-empirical predictions of \cite{Ellison_1957}, which are not apparently consistent with these flows. In particular, Ellison argued, through consideration of the turbulent energy equations, that the turbulent Prandtl number, $Pr_T$ diverges to infinity as $Ri_f \rightarrow  0.15$. This is significant, as $Pr_T$ and $Ri_f$ can be directly related through an appropriately defined Richardson number, and if, as Ellison assumed, `turbulence can be maintained at large values' of this Richardson number, then the picture is self-consistent, and such `strongly stratified' turbulence would be expected to exhibit this particular value of the flux Richardson number. Of course, the key question is how the various quantities, in particular the turbulent Prandtl number and the Richardson number, are actually defined, as the gradient Richardson number $Ri_g(z,t)$ is inevitably a function of space and time.

\begin{figure}
  \centering 
  \includegraphics[width=\textwidth]{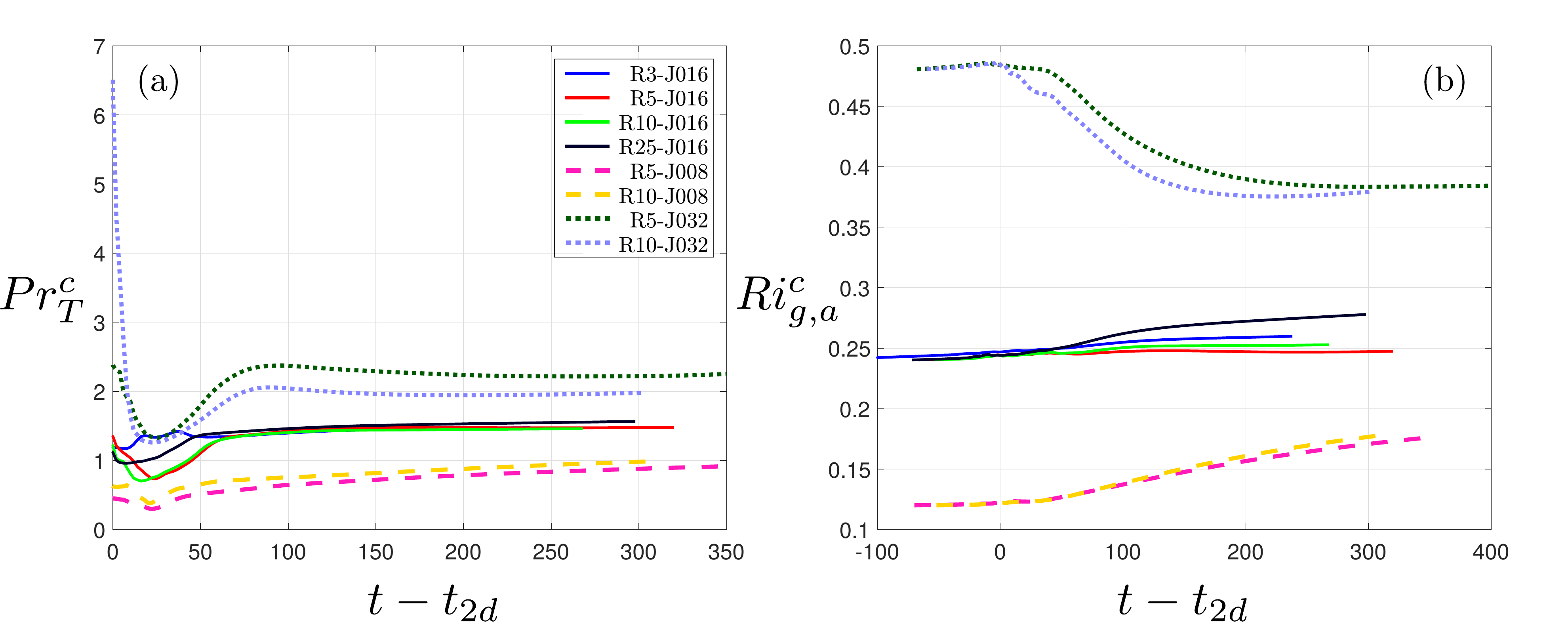}
  \caption{Variation with time of (a) the cumulative turbulent Prandtl number, $Pr_T^c$ and (b) the cumulative averaged gradient Richardson number $Ri^c_{g,a}$ as defined in \eqref{eq:Prt}.}
    \label{fig:cum_Prt}%
\end{figure}

To investigate this issue further, we define a cumulative turbulent Prandtl number, $Pr^c_T = K^c_m/K^c_\rho$ based on the cumulative (and crucially irreversible) diapycnal diffusivity of mass (i.e. $K^c_\rho$) and momentum (i.e. $K^c_m$) which are defined as
\begin{equation}
K^c_\rho(t) = \frac{\mathcal{M}_c(t)}{N^2_c(t)}, \qquad
K^c_m(t)    = \frac{\mathcal{M}_c(t)+\mathcal{D}_c(t)}{S^2_c(t)},  
\label{eq:diapycnal kappa}
\end{equation}
following \cite{SP15}. In these definitions, the denominators are defined as appropriately time-averaged and vertically-averaged
squares of the buoyancy frequency and shear:
\begin{equation}
N_c^2 (t) = \frac{1}{t} \int_0^t \langle N^2 \rangle \, dt, 
\qquad 
S^2_c = \frac{1}{t}\int_0^t \langle S^2 \rangle \, dt , 
\label{eq:ncscdef}
\end{equation}
in which $N^2(z,t)$ and $S^2(z,t)$ are defined in \eqref{eq:Rig}. Hence the cumulative turbulent Prandtl number $Pr^c_T$ can be related to the cumulative mixing efficiency, and an appropriate `cumulative averaged gradient' Richardson number $Ri_{g,a}^c(t)$  as 
\begin{equation}
Pr^c_T (t) = \frac{K^c_m(t)}{K^c_\rho(t)} = \frac{Ri_{g,a}^c(t)}{\mathcal{E}_c(t)} ,
\qquad Ri_{g,a} ^c (t) = \frac{N_c^2(t)}{S_c^2(t)} .
\label{eq:Prt}
\end{equation}
When averaged over the entire life-cycle of a turbulent event,  \eqref{eq:Prt} converges to
\begin{equation}
\widetilde{Pr}_T = \frac{\widetilde{Ri}_{g,a}}{\widetilde{Ri}_f}. 
\label{eq:rifprt}
\end{equation}
Expressed in this way, it is now clear what must be meant in this context for `turbulence to be maintained' at strong stratification. Since $\widetilde{Ri}_f < 1$ by construction, $\widetilde{Pr}_T \rightarrow \infty$  requires turbulence to remain at large values of $\widetilde{Ri}_{g,a}$. 

Figure \ref{fig:cum_Prt} illustrates $Pr_T^c(t)$ and $Ri_{g,a}^c(t)$ for all DNS analyses of HWI investigated herein. The corresponding life-cycle-averaged quantities, i.e. $\widetilde{Pr}_T$ and $\widetilde{Ri}_{g,a}$ should be realized in these plots as the ultimate values of $Pr_T^c(t)$ and $Ri_{g,a}^c(t)$ as $t \to \tau$. Obviously for the investigated range of the initial bulk Richardson numbers, $\widetilde{Pr}_T$ does not demonstrate a single universal number, in the sense of that observed for $\widetilde{\Gamma}$ in figure \ref{fig:cum_soc}(b), but nevertheless, $\widetilde{Pr}_T$ remains of $O(1)$ while $\widetilde{\mathcal{E}} = \widetilde{Ri}_f \sim 1/6$ in apparent contradiction with the semi-empirical expression derived by \cite{Ellison_1957}.

Furthermore, it is apparent in figure \ref{fig:cum_Prt}(b) that $Ri_{g,a}^c(t)$ is typically larger than the bulk Richardson number $Ri_b$ (as defined in \eqref{eq:Rib}) for any particular flow, and furthermore  $Ri_{g,a}^c(t)$  increases with increasing $Ri_b$. Once again there is no evidence of `universal' behavior for this vertically averaged cumulative quantity, unlike the behavior of the (inherently local and time-dependent) gradient Richardson number $Ri_g(z,t)$. Since, as shown in figure \ref{fig:cum_soc}(a) $\mathcal{E}_c(t) \rightarrow 1/6$ quite rapidly, this observation also implies (consistently with the data shown in figure \ref{fig:cum_Prt}(a)) that $\widetilde{Pr}_T$ increases with $Ri_b$. Since there is predicted to be a range of wavenumbers susceptible to primary HWI as $Ri_b$ increases to arbitrarily large values (albeit this range narrows as $Ri_b$ increases) it is thus possible that $\widetilde{Pr}_T \rightarrow \infty$ as $Ri_b$ and hence $\widetilde{Ri}_{g,a}$ tends to very large values, with HWI-induced turbulence still occurring, although we have not been able to investigate such truly `large' values of $Ri_b$.

However, it is very important to appreciate that, even if this behavior occurs, our observations are not consistent with Ellison's model, since within his model the flux Richardson number approaching $0.15$  \emph{implies} that the turbulent Prandtl number tends to infinity, whereas we find that $\widetilde{Ri}_f \simeq 0.16$ while $\widetilde{Pr}_T \sim O(1)$. Crucially, where the flow is turbulent, the appropriate measure of the stratification $Ri_g(z,t)$ \emph{self-organises} to a critical value close to $1/4$ irrespective of the external parameters, suggesting that strongly stratified \emph{local} values are not accessible to such unforced shear layers. This observation calls into question whether it is ever appropriate to consider such flows 
as being `strongly stratified', even when the bulk Richardson number $Ri_b$ has a relatively large value, except perhaps in the sense that flows susceptible to primary HWI are characterised by high values of the gradient Richardson number initially in the immediate vicinity of the density interface.

\subsection{HWI-induced turbulence and scale-invariance}
\label{sec:1/f}

The concept of self-organized criticality was originally proposed to explain ubiquitous observations of scale invariance as characterized by power laws of the form $1/f^{\beta}$ with $\beta \sim ~1$ \citep{BTW}. Such power laws are often observed for the size, strength, duration or number of the `avalanches' in slowly driven complex systems, as in the problem of turbulent thermal convection with an endothermic phase transition within the layer mentioned previously. Nonetheless, as noted by \cite{SOC_book_2012}, the scale invariant state has often been misconstrued as being synonymous with SOC although scale invariance is actually a consequence of SOC. 

This issue becomes relevant to our discussion in the current paper because, as we have previously shown \citep{SCP15}, the streamwise kinetic energy associated with both KHI- and HWI-induced turbulence exhibits a $-5/3$ power law for a range of scales sufficiently larger than the dissipative subrange, and specifically above the Ozmidov scale (as defined in \eqref{eq:Reb}) whereas as discussed in this paper, only the turbulence induced by slowly evolving HWI is self-organized into a critical state.

\begin{figure}
  \centering 
  \includegraphics[width=\textwidth]{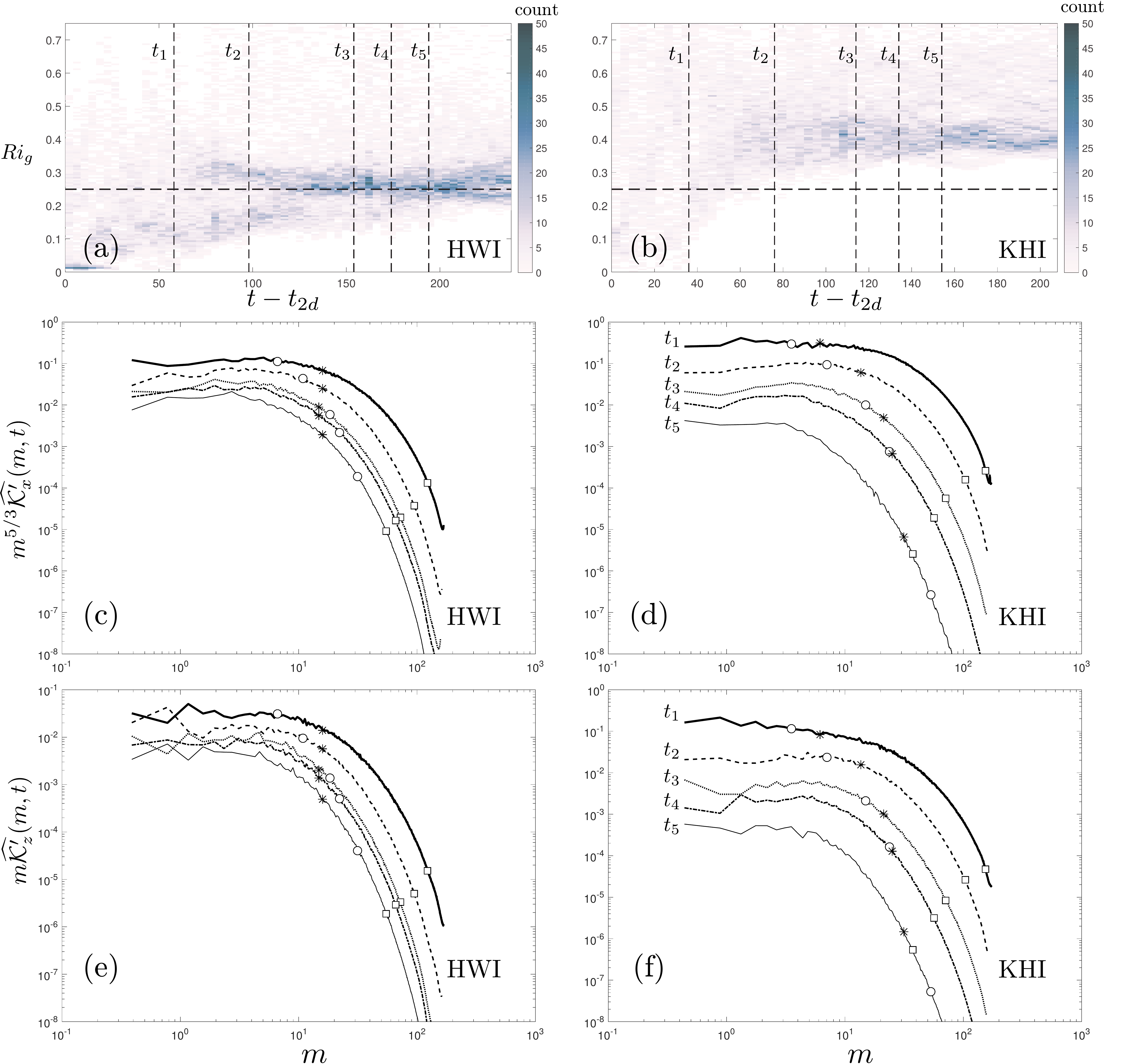}
  \caption{ 
Top row: Two-dimensional histograms of $Ri_g(z,t)$ as a function of time for $t\geq t_{2d}$;
Middle row: streamwise spectra of $\widehat{\mathcal{K}_x}(m, t)$ (as defined in \eqref{eq:spectraKx}) compensated with $m^{5/3}$;
Bottom row: streamwise spectra of $\widehat{\mathcal{K}_z}(m, t)$ (as defined \eqref{eq:spectraKz}) compensated with $m$;
for the case R3-J016 with HWI-induced turbulence (left column) and the case with KHI-induced turbulence (right column). The spectra are calculated at times $t_1=t_{3d}$, $t_2 = t_{3d} + 40$, $t_3 = t_{rl}$, $t_4 = t_{rl} + 40$ and $t_5 = t_{rl} + 80$. These times are indicated by vertical dashed lines in (a,b)  whose corresponding spectra are appropriately distinguished by a specific line style as labeled in (d, f). Also on these spectra, at each time and for each simulation, the wavenumber scales corresponding to energy-containing length scales, $\ell_{en}$ (see \eqref{eq:L_en}) as well as the Ozmidov ($\ell_O$) and Kolmogorov ($\ell_K$) length scales (see \eqref{eq:Reb}) are marked respectively by $\ast$, $\circ$ and $\Box$ and are computed as $1/\ell$.}
    \label{fig:1/f}%
\end{figure}

To illustrate this issue further, we investigate the power spectral density of streamwise and vertical perturbation velocities as captured respectively by the streamwise spectra of streamwise and vertical perturbation kinetic energy defined as
\begin{eqnarray}
 \widehat{\mathcal{K}'}_x(m,t) &=& \frac{\pi}{L_y} \sum_{n} \langle \widehat{u'} \widehat{u'}^* \rangle
 \label{eq:spectraKx}
 \\
 \widehat{\mathcal{K}'}_z(m,t) &=& \frac{\pi}{L_y} \sum_{n} \langle \widehat{\,w'\,} \widehat{\,w'\,}^* \rangle
 \label{eq:spectraKz}
\end{eqnarray}
in which $\widehat{\boldsymbol{u}'} = (\widehat{u'}, \widehat{v'}, \widehat{w'})$ represents the horizontal two-dimensional Fourier transform (denoted by a hat) of the perturbation velocity field $\boldsymbol{u}'$ defined in \eqref{eq:u=U+u'}. The asterisks denote complex conjugation and, as usual, $\langle . \rangle$ denotes vertical averaging and $(m,n)$ are the streamwise and spanwise wavenumbers.  

%

In figure \ref{fig:1/f}, we plot $\widehat{\mathcal{K}'}_x(m,t)$ (middle row) and $\widehat{\mathcal{K}'}_z(m,t)$ (bottom row) in compensated forms at five different times during the flow evolution of a specific case of HWI-induced turbulence (case R3-J016, left column) and the flow with KHI-induced turbulence (right column). It is important to remember that these two cases only differ in terms of their initial values of $R$ (see table \ref{tab:siminfo}). These five times are indicated on the two-dimensional histograms of $Ri_g(z,t)$ plotted in the top row of figure \ref{fig:1/f}, and are associated with $t_1=t_{3d}$, $t_2 = t_{3d} + 40$, $t_3 = t_{rl}$, $t_4 = t_{rl} + 40$ and $t_5 = t_{rl} + 80$. The corresponding spectra at each of these times is distinguished by a distinct line style as labeled. 

For both KHI-induced turbulence and HWI-induced turbulence, the streamwise spectra of $\widehat{\mathcal{K}'}_x$ and $\widehat{\mathcal{K}'}_z$ exhibit $m^{-5/3}$ and $m^{-1}$ power laws for scales that are sufficiently larger than the Ozmidov length scale ($\ell_O$, indicated on the spectra by a circle). This indicates that the horizontal component of the motion decays more rapidly with wavenumber than does the vertical component, which in turn implies oblate flattened structures (as expected in vertically stratified flows) that are self-similar across a range of wave numbers and hence are scale invariant. Such observed scale-invariance appears to hold at all the investigated characteristic times. As a result, despite the inherent difference between KHI- and HWI-induced turbulence in terms of the self-regulation of the HWI-induced turbulence towards a critical state, they are both characterized by a scale-invariance in their inertial sub-range. This implies again that scale-invariance is a necessary but \emph{not} a sufficient condition for the emergence of SOC.

The energy-containing scale, $\ell_{en}$, that might reasonably be associated with the smallest occurring `avalanches' may be defined as,
\begin{equation}
 \ell_{en} = \alpha^2 \left(\frac{\mathcal{Q}^3}{\mathcal{D}} \right),
 \label{eq:L_en}
\end{equation}
in which $\mathcal{Q}^2 = (2/3)\mathcal{K}'$, and where once again the scale factor $\alpha = (\ell_u/L_z)^{1/4}$ has been introduced to remove the scale dependence of the vertical averaging on computational domain size. 

\begin{figure}
  \centering 
  \includegraphics[width=\textwidth]{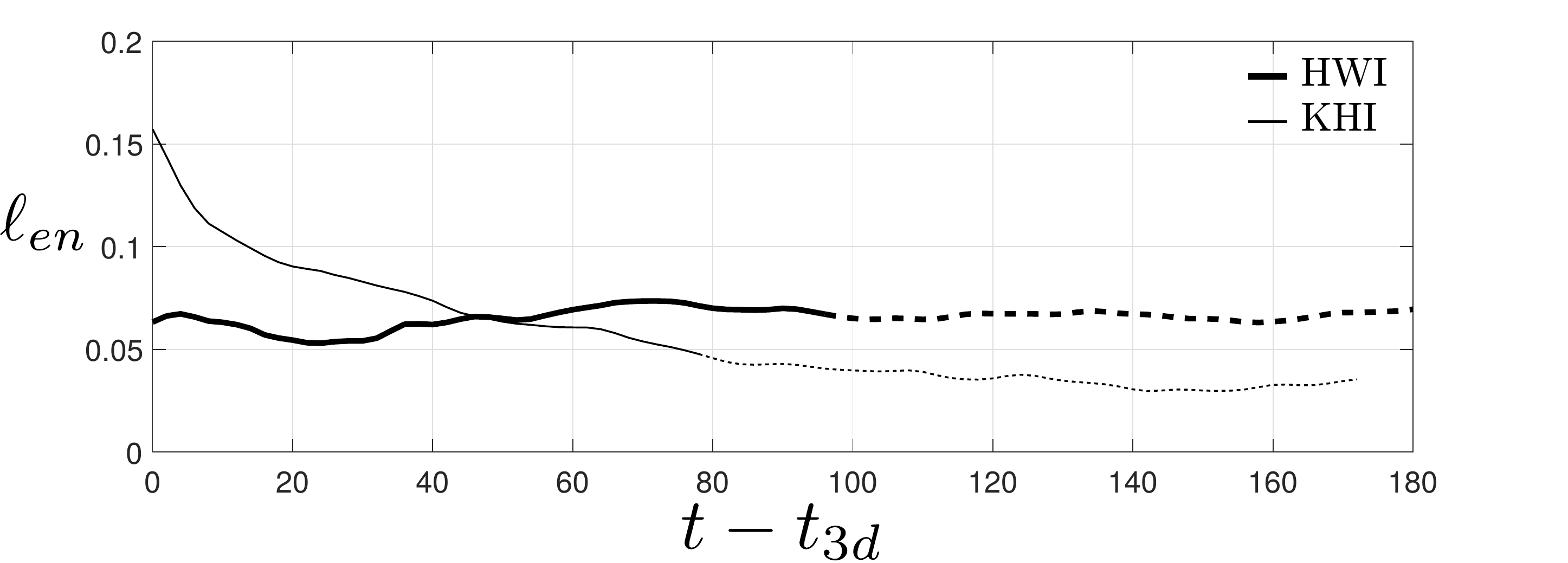}
  \caption{The variation with time of $\ell_{en}$ after the onset of fully three-dimensional flow (i.e. $t\geq t_{3d}$) associated with the HWI-induced turbulence (case R3-J016, shown by thicker lines) and the KHI-induced turbulence (shown by thinner lines). The re-laminarized phase beginning at $t_{rl}$ is indicated by dashed lines.}
    \label{fig:L_en}%
\end{figure}

Figure \ref{fig:L_en} illustrates the evolution of $\ell_{en}$ for the same two DNS cases analysed in figure \ref{fig:1/f}. As also demonstrated in figure \ref{fig:L_en} by $\ast$, the energy-containing scale $\ell_{en}$, noticeably goes to higher wavenumber (smaller scale) for the KHI-induced turbulence, while it remains much closer to constant in the flow characterised by HWI-induced turbulence. This time-invariance of $\ell_{en}$ also appears to be a characteristic behavior of HWI-induced turbulence, giving further support to the argument that HWI-induced turbulence is in `some kind of equilibrium' (c.f. figure \ref{fig:F}).

As observed in figure \ref{fig:1/f}, the energy-containing length scales that are $\gtrsim \mathcal{O}(\ell_{en})$ appear to represent the scale-invariant `avalanches'. Nevertheless, it is not immediately obvious what the most appropriate way is to measure the respective size of such individual `avalanches' that are widespread and localized in these turbulent flows. We may alternatively demonstrate the scale-invariance of such `avalanches' in physical space by introducing a point-wise measure of the energy-containing length scale, $L_{en}(\boldsymbol{x},t)$, as,
\begin{equation}
 L_{en}(\boldsymbol{x},t) = \frac{q^3(\boldsymbol{x},t)}{\epsilon'(\boldsymbol{x},t)},
 \label{eq:Len(x,t)}
\end{equation}
in which, similar to their volume-averaged counterparts employed in \eqref{eq:L_en}, $q^2 = (2/3)k'(\boldsymbol{x},t)$ where $k'(\boldsymbol{x},t) = 0.5 (\boldsymbol{u}'\cdot\boldsymbol{u}')$ is the point-wise turbulent kinetic energy (i.e. $\mathcal{K}' = \langle \overline{k'} \rangle$).

\begin{figure}
  \centering 
  \includegraphics[width=.75\textwidth]{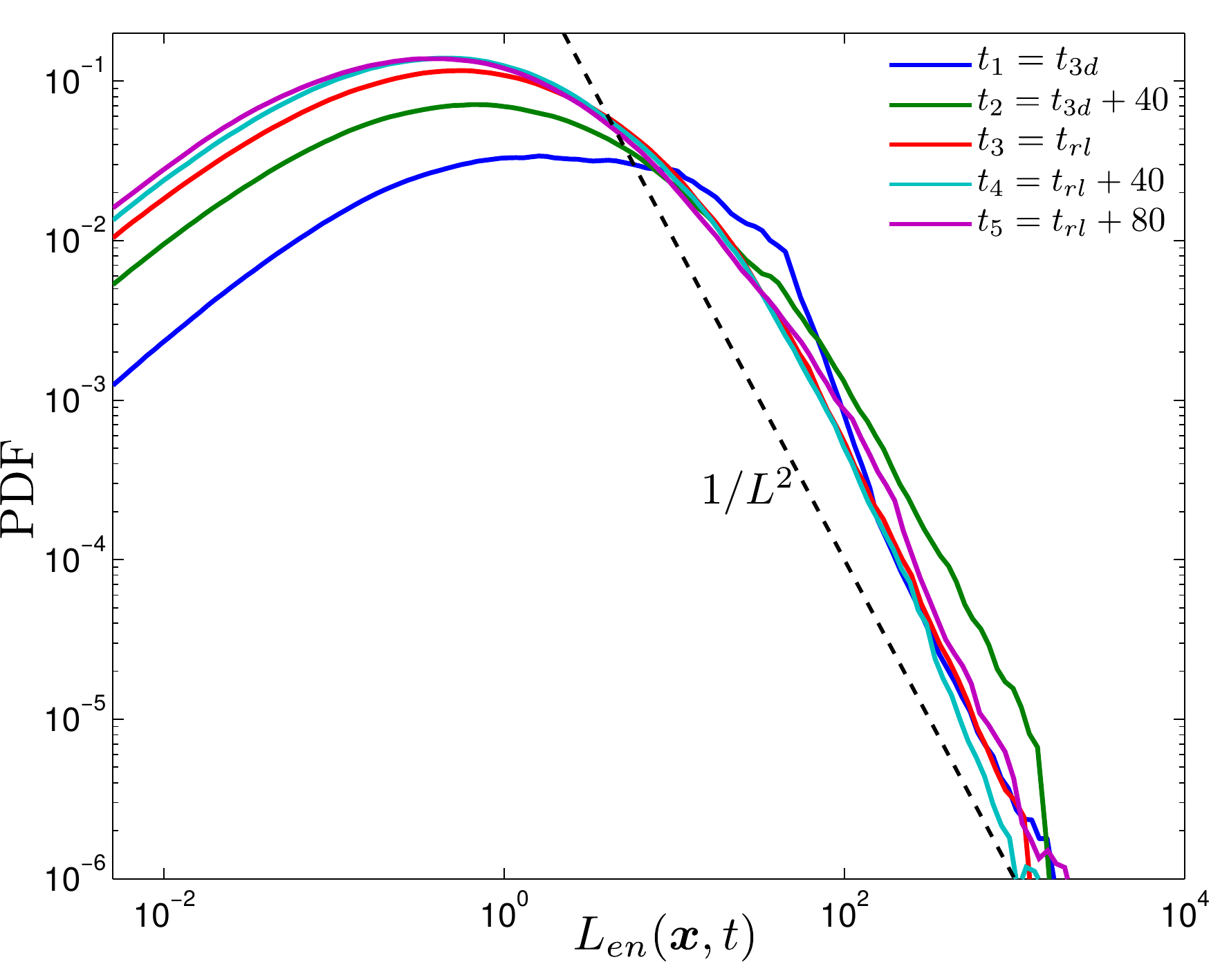}
  \caption{The probability density function (PDF) of $L_{en}(\boldsymbol{x},t)$ as defined in \eqref{eq:Len(x,t)} at five characteristic times (as denoted in the legend) for case R3-J016. The binning has been conducted for data within a three-dimensional box with the same horizontal extent as the computational domain but limited vertically to $-2.5 \leq z\leq 2.5$. Also bins with less than five members have been discarded.}
    \label{fig:scale_invariance}%
\end{figure}

Figure \ref{fig:scale_invariance} illustrates the probability density functions of the estimated local energy-containing scales, $\mbox{PDF}(L_{en})$, for the HWI-induced turbulence (case R3-J016) at the same five characteristic times discussed in figure \ref{fig:1/f}. To eliminate unstratified, non-turbulent regions from our analysis the PDFs are associated with a confined three-dimensional box with its non-dimensional height limited to $-2.5 \leq z\leq 2.5$. The roll-off at small scales in these PDFs is associated with the finite resolution of the DNS analysis.

A striking power-law distribution of the form $\sim L_{en}^{-2}$ is evident for all the times investigated in figure \ref{fig:scale_invariance} which suggests an interesting self-similar and scale-invariant relationship between the local turbulent kinetic energy ($k'$) and the local turbulent dissipation rate ($\epsilon'$). In other words, the localized distribution of $k'$ and $\epsilon'$ are self-organized (consistent with our previous discussions concerning figure \ref{fig:SOC_eps}) in such a way that the \emph{implied} localized energy-containing length scale, $L_{en}$, follows a power-law distribution. Notice that, relevant to our discussion here is only the \emph{distribution} of $L_{en}$ (i.e. $\mbox{PDF}(L_{en})$, \emph{not} $L_{en}$) that demonstrates a scale-free relationship between $k'$ and $\epsilon'$. In fact, the characteristic length scales of turbulence are only meaningful in a statistical `bulk' sense after a spatial or temporal filtering or averaging operator is employed to identify a characteristic time scale and a characteristic velocity scale. Therefore it is inappropriate to make any quantitative comparison between specific values of $\ell_K$, $\ell_O$ and $\ell_{en}$ as presented in figure \ref{fig:1/f} and those associated with the point-wise values of $L_{en}$ which are indeed orders of magnitude larger.

\section{Conclusions}
\label{sec:summary}
%

Classic arguments of \cite{Turner_1979} based on the Monin-Obukhov similarity theory for wall-bounded shear flows suggest a `self-regulated' state under `very stable' conditions which leads to constant values for the gradient Richardson number and mixing efficiency. In a different branch of science, in the context of highly interacting complex dynamical systems, the notion of `self-organized criticality' (SOC) has been identified as the underlying mechanism of many non-equilibrium slowly-driven systems that reveal a scale-invariant state. In addition, there has been accumulating evidence based on oceanic observations of turbulent flows for the ubiquitous measurement of a gradient Richardson number $Ri_g \sim 1/4$ that is curiously close to its critical value based on inviscid linear stability theory \citep{Smyth_etal_2013_marginal, Holleman_Geyer_2016}.

Motivated by these ideas, we have investigated the various characteristic properties of the turbulent states of (relatively) strongly stratified freely evolving shear flows. We have employed direct numerical simulations to study Holmboe wave instability (HWI). Crucially, HWI is a type of shear instability, which unlike its better known relative, the Kelvin-Helmholtz instability (KHI), occurs for arbitrarily large values of the gradient Richardson number at the interface (i.e. $Ri_g(0)$) and arbitrarily large values of the bulk Richardson number $Ri_b$, provided that the density layer is significantly sharper than the shear layer.  Unlike KHI, which is suppressed for shear layer midpoint gradient Richardson numbers $Ri_g(0) > 1/4$, under such strongly stratified conditions (in the sense here of both high values of gradient Richardson number in the vicinity of the density interface and sufficiently high values of $Ri_b$) HWI not only emerges but can also induce highly turbulent flows at sufficiently high Reynolds numbers. 

A key distinguishing characteristic between flows susceptible to primary KHI and flows susceptible to primary HWI relates to the longevity of the associated induced turbulence. The localization of `scouring' motions near the interface for HWI-induced turbulence  manifests itself energetically by inducing turbulent flows that evolve towards a state of `quasi-equilibrium'. In this state, the decay rate of total available turbulent energy, consisting of turbulent kinetic energy and available potential energy, is only weakly time-dependent and therefore the turbulence is sustained and long-lived.  In fact, this `quasi-equilibrium' state is achieved independently of the initial conditions for HWI. In particular, the quasi-equilibrium state is independent of the strength of the initial bulk stratification. The KHI-induced turbulence, on the other hand, can be grossly out of equilibrium and hence short-lived, due to the strong imprint or memory of large-scale KHI overturns which substantially stir the fluid leading to a rapid rise in its available potential energy. For these short-lived turbulent flows induced by KHI, the critical state of marginal instability therefore appears to be irrelevant. Again to invoke the sandpile analogy, a sandpile is unlikely to be formed if sand is loaded onto the pile suddenly, while conversely a slow addition of sand grains does indeed lead to maintenance of the critical slope through the continual occurrence of avalanches.

\begin{figure}
    \centerline{\includegraphics[width=0.55\textwidth]{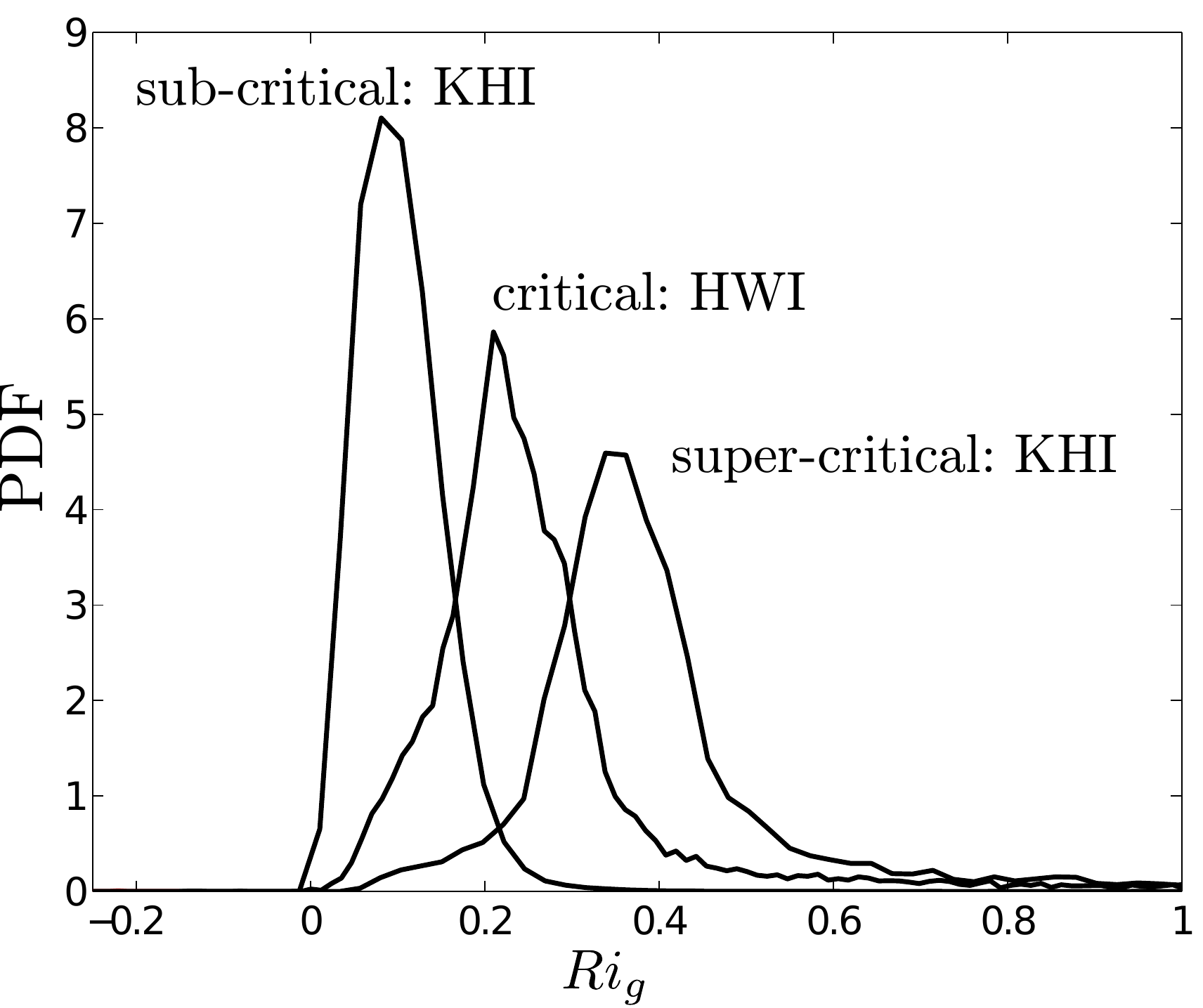}}
    \caption{Probability density functions for $Ri_g$ for a typical `critical' case of HWI-induced turbulence (case R10-J016); for a supercritical case of KHI-induced turbulence (the flow susceptible to KHI listed in table \ref{tab:siminfo}); and for a subcritical case of KHI-induced turbulence, corresponding to a flow with $Ri_b=0.04$ and the same other parameters as the other two cases, previously described in \cite{SP15} (see their table 1).}
    \label{fig:PDF_HWI_KHI}%
\end{figure}

Furthermore, we have demonstrated that HWI-induced turbulence organizes itself, prior to re-laminarization, towards a critical state that is characterized by a horizontally-averaged mean flow with a probability density function strongly concentrated in the near vicinity of $Ri_g \sim 1/4$. This so-called self-organised criticality (SOC) appears to be independent of the initial conditions and hence requires no external tuning. Thus \emph{universal} behavior is expected for strongly stratified turbulent shear flows induced by HWI.

While for HWI-induced turbulence, $Ri_g(z,t)$ is attracted towards the \emph{critical} value of $Ri_g\sim 1/4$, for KHI-induced turbulence, the PDFs of $Ri_g(z,t)$ point towards (in general) either \emph{sub-critical} or \emph{super-critical} states for the most-likely or `peak' value of $Ri_g$, which we define as the peak value of the PDF being either smaller than or greater than $1/4$ respectively. This important qualitative difference is plotted in figure \ref{fig:PDF_HWI_KHI} (c.f. figure \ref{fig:Rig_PDF}). There is a unique initial choice of $Ri_b$ for flows susceptible to KHI to reach the critical state, and so
flows susceptible to primary KHI require external tuning (through changing the initial conditions of the bulk stratification to just the right value) to reach the critical state while flows susceptible to HWI robustly self-organize towards this critical state largely independently of the chosen initial conditions.

Perhaps most interestingly, we have demonstrated a characteristic and apparently universal cumulative partitioning of energy over the entire life-cycle of flows susceptible to primary HWI, with the total cumulative irreversible mixing appearing to be always approximately $20\%$ of the cumulative turbulent kinetic energy dissipation due to molecular viscosity. This implies that an `Osborn-like' turbulent flux coefficient of $\Gamma_c \sim 0.2$ can be associated with  turbulent flows in a generic self-organized critical state characterized by local values of $Ri_g \simeq 1/4$, arising from the break down of Holmboe wave instabilities at sufficiently high $Re$ in stratified shear flows with relatively sharp initial density interfaces.

%
\section{Acknowledgement}
\label{sec:acknowledgement}
H.S. acknowledges the SOSCIP TalentEdge post doctoral fellowship and is grateful to the David Crighton Fellowship from D.A.M.T.P., University of Cambridge. All the computations were performed on the BG/Q supercomputer at the University of Toronto which is operated by SciNet for the Southern Ontario Smart Computing Innovation Platform. SciNet is funded by: the Canada Foundation for Innovation under the auspices of Compute Canada; the Government of Ontario; Ontario Research Fund - Research Excellence; and the University of Toronto. The research of W.R.P. at the University of Toronto is sponsored by NSERC Discovery Grant A9627. The research activity of C.P.C. is supported by EPSRC Programme Grant EP/K034529/1 entitled `Mathematical Underpinning of Stratified Turbulence'. All codes and initial conditions used to generate the data used in this paper, in particular to construct the figures, are available at \url{
https://github.com/hsalehipour/mixing_analysis}.

\bibliographystyle{jfm}
\bibliography{SOC}

\end{document}